\documentclass{aa}
\usepackage[dvips]{graphicx}
\newcommand{\asec}{$^{\prime\prime}$}
\newcommand{\pam}{.\hskip-2pt$^{\prime}$}
\newcommand{\pas}{.\hskip-2pt$^{\prime\prime}$}
\def\MI{I21307}
\def\MII{I22172}
\def\AMM{NH$_3$}
\def\ace{CH$_{3}$C$_{2}$H}
\def\CI{\mbox{$^{13}$CO}}
\def\CII{\mbox{C$^{18}$O}}

\def\CO{\mbox{$^{12}$CO}}
\def\HCOp{\mbox{HCO$^+$}}
\def\HCOpI{\mbox{H$^{13}$CO$^+$}}

\def\HII{H{\sc ii}}

\def\kms{\mbox{km~s$^{-1}$}}

\def\Tmb{\mbox{$T_{\rm MB}$}}

\begin{document}
\title{Nature of two massive protostellar candidates: \mbox{IRAS 21307+5049} and \mbox{IRAS 22172+5549} }
\author{F. Fontani \inst{1} \and R. Cesaroni \inst{2}
	\and L. Testi \inst{2} 
	\and S. Molinari \inst{3}  \and Q. Zhang \inst{4}
	\and J. Brand \inst{5}
	\and C.M. Walmsley \inst {2}
        }
\institute{Dipartimento di Astronomia e Fisica dello spazio, Largo E. Fermi 2,
           I-50125 Firenze, Italy \and
	   INAF, Osservatorio Astrofisico di Arcetri, Largo E. Fermi 5,
           I-50125 Firenze, Italy \and
	   IFSI, CNR, Via Fosso del Cavaliere, I-00133 Roma, Italy \and
	   Harvard Smithsonian Center for Astrophysics, 60 Garden Street,
           Cambridge, MA 02138 \and
	   Istituto di Radioastronomia, CNR, Via Gobetti 101, I-40129 Bologna,
	   Italy 
 }
\offprints{F. Fontani, \email{fontani@arcetri.astro.it}}
\date{Received date; accepted date}

\titlerunning{IRAS 21307+5049 and IRAS 22172+5549}
\authorrunning{Fontani et al.}

\abstract{We present observations of continuum and molecular lines towards the
protostar candidates IRAS 21307+5049 and IRAS 22172+5549. 
Single-dish
maps in the \CO\ (2--1), \CII\ (2--1), \HCOpI\ (1--0) lines and 850 $\mu$m 
continuum are compared with interferometric maps in the \CO\ (1--0) line and
3~mm continuum, and with mid- and near-infrared images. Also, single-pointing
spectra of the \ace\ (6--5), (8--7) and (13--12) lines observed towards
IRAS 21307+5049, and H$_{2}$ and [FeII] line emission observed towards
IRAS 22172+5549, are presented. A plausible interpretation of our data
based on the continuum maps and spectral energy distributions is that
two components are present: a compact molecular core, responsible for the 
continuum emission
at wavelengths longer than $\sim25\;\mu$m, and a cluster of stars 
located close to the center of the core, but not spatially coincident
with it, responsible for the emission at shorter wavelengths. 
The core is approximately located at the center of the 
associated molecular outflow, detected for both sources in the \CO\ (1--0) and 
(2--1) lines. The cores have masses of $\sim50\;M_{\odot}$,
and luminosities of $\sim10^{3}\;L_{\odot}$. The outflows parameters 
are consistent with those typically found in high-mass
young stellar objects. 
Our results support the hypothesis that in these sources the luminosity 
is dominated by accretion rather than by nuclear burning.
We conclude that the sources embedded inside the cores are likely 
protostars with mass $\sim 5-8\;M_{\odot}$.
}

\maketitle

\section{Introduction}

Molecular outflows are a well known phenomenon associated with the star
formation process. Their origin is related to the
formation of proto-stellar disks, and they play an important role in the
accretion phase of a protostar. In fact, they
increase the turbulence of the surrounding material, hence contributing 
to support the molecular cloud against gravitational
contraction and perhaps halting the collapse onto the forming star.
The role of outflows is especially relevant in the context of the 
{\it high-mass}
($M\geq 8\;{\rm M_{\odot}}$) star formation process. While low-mass 
stars are believed to form through accretion, the mechanism for the 
formation of high-mass stars is still unclear. Two scenarios have been 
proposed: accretion (Behrend \& Maeder \cite{behrend}), or 
coalescence of several lower mass protostars
(Bonnell et al. \cite{bonnell}; Stahler et al. \cite{stahler}). Theoretical 
models predict that
the accretion process naturally leads to the formation of disks and 
collimated bipolar 
outflows driven by the forming star, whereas in the coalescence scenario the
outflows must be poorly collimated: it is thus important to 
investigate at high angular resolution the presence
of collimated outflows originating from massive protostellar candidates
in order to discriminate between coalescence and accretion.

Although many studies have been carried out on outflows originating from
low-mass objects, only a limited number of observations are available for
those associated with high-mass young stellar objects (YSOs). Moreover,
most of these have been performed at low angular resolution (e.g. Zhang et al. 
\cite{zhang}; Beuther et al. \cite{beuther}). 

In this paper, we investigate the structure of two candidate 
massive protostars: IRAS 21307+5049 (hereafter \MI ) and IRAS 22172+5549
(hereafter \MII ). These sources,
located at a kinematic distance of 3.2 and 2.4 kpc respectively (Molinari et
al. \cite{mol02}), are part of an initial large sample
of luminous IRAS sources. In a series of studies, Molinari et al. 
(\cite{mol96}, \cite{mol98a}, \cite{mol98b}, \cite{mol00}) selected from the
initial sample of 260 sources, a small number of likely massive YSOs 
thought to be in
an early evolutionary stage preceding the formation of a compact \HII\
region. Then, more recently, Molinari et al. (\cite{mol02}) observed
11 of these sources at high angular resolution with the Owens Valley Radio 
Observatory (OVRO) in the \HCOp\ (1--0) line and in the millimeter continuum,
and in the 3.6 cm radio continuum with the Very Large Array (VLA). For 
several of
these sources, a compact molecular core was detected, and in several
cases this core does not show emission at 3.6~cm. Moreover, in a few cases,
non gaussian wings were identified in the \HCOp\ (1--0) line, suggesting the
presence of outflows.  
These results
indicate that in the OVRO sample there is a number of sources which
present observational features expected for protostars.
This is confirmed by Fontani et al. 
(\cite{fonta2}) who observed one of these sources, IRAS 23385+6053,
with single dish telescopes and interferometers at various 
wavelengths: the authors have demonstrated the presence of a massive 
dusty core, which likely hosts
a young stellar object in the protostellar phase.

Two other sources of the OVRO sample, 
\MI\ and \MII , show the presence of a compact dusty
core without centimeter continuum emission, with prominent wings in the
\HCOp\ (1--0) line, and characterised by IRAS luminosities of a few
$10^{3}\;L_{\odot}$
(Molinari et al. \cite{mol02}). Hence, they represent likely massive YSOs, 
in an evolutionary stage preceding the formation of
a compact \HII\ region. Recently, Zhang et al. (\cite{zhang}) and 
Molinari et al. (\cite{mol02}) have detected towards both sources a 
compact outflow in the \CO\ (2--1), \CI\ (2--1) and \HCOp\ (1--0) lines. 
Here we present the results obtained from single-dish and 
interferometric observations towards \MI\ and \MII , aimed at 
investigating their nature.

In Sect.~\ref{obs} we describe the observations and the
methods used for data reduction. In Sect.~\ref{res}, we present the 
physical parameters of the sources and associated 
outflows. In Sect.~\ref{discu}, we discuss our
findings, and give a summary and the main conclusions in Sect.~\ref{conc}.

\section{Observations and data reduction}
\label{obs}
The main characteristics of the observed sources are given in 
Table~\ref{tsources}.
The coordinates corresponding to the sub-mm peak position 
(Molinari et al. \cite{mol00}) in \MI\ and to the 3.4~mm emission 
(Molinari et al. \cite{mol02}) in \MII , are listed
in Cols.~2 and 3; the kinematic distance, $d$, and the Galactocentric distance,
$R_{\rm GC}$,
are given in Cols. 4 and 5, while the bolometric luminosity, $L$ (from Molinari
et al. \cite{mol02}), and the source 
velocity, $v_{\rm LSR}$ ($v_{\rm NH_{3}}$ from Molinari et al. 
\cite{mol96}), are shown in Cols.~5 and 6, respectively.

Table~\ref{tobs} lists all molecular transitions observed. For each line,
we give the rest frequency (Col.~2), the telescopes used (Col.~3), the half 
power beam width (HPBW) (Col.~4), the total bandwidth (Col.~5) and the 
spectral resolution (Col.~6).

\subsection{IRAM-30m observations}
\label{30m}

\subsubsection{\CO\ (2--1) and \CII\ (2--1)}

The observations of the \CO\ (2--1) line were carried out on February 28 
2002 towards \MI , and July 14 2003 towards \MII . Those of the \CII\ (2--1)
line were performed on July 13 2003.
We made $5^{\prime}\times 5^{\prime}$ maps in the \CO\ (2--1) line, 
and $2^{\prime}\times 2^{\prime}$ maps in the \CII\ (2--1) line using 
the HERA
mutibeam receiver in the on-the-fly mode, in which the telescope moves across
the source and dumps the data with 4$^{\prime\prime}$ sampling. The dump
time was 2 sec per position. We
used the receiver with a derotator angle of 9.6$^{\circ}$. 
The observations 
were made in position switching mode for the \CO\ (2--1) line, and
in frequency switching mode for the \CII\ (2--1) line. 
The reference positions used for the \CO\ (2--1) maps were
$\alpha({\rm J}2000)=21^{h} 55^{m} $28\fs 0 and $\delta({\rm J}2000)=48^{\circ} 10^{\prime} 37^{\prime\prime}$ for \MI , and $\alpha({\rm J}2000)=22^{h} 25^{m} 55$\fs 8 and 
$\delta({\rm J}2000)=54^{\circ} 35^{\prime} 07^{\prime\prime}$ for \MII , and
they are CO-free at levels of $\sim 0.2$ K.
Telescope HPBW, total bandwidth and spectral resolution are listed in Table
\ref{tobs}.
Focus and pointing
were checked every hour by means of continuum cross scans on Uranus or
1253-055. The data were calibrated with the
``chopper wheel'' technique described in Kutner \& Ulich (\cite{kutner}).
Calibration was checked by measuring the \CO\ (2--1) line intensity towards
the ultracompact \HII\ region W51D. 

\subsubsection{ \ace , \HCOpI\ and SiO}

The \ace , \HCOpI\ and SiO lines were observed towards \MII\
on August 11, 1999. 

For the \ace\ lines, we simultaneously observed the $J=$(6--5), (8--7) and
(13--12) rotational transitions in the 3~mm, 2~mm and 1.3~mm bands,
respectively. The observations were made towards the nominal central position
of \MII\ given in Table~\ref{tsources}. We used two spectrometers
simultaneously: an autocorrelator, which covered the lowest $K$ components
with high spectral resolution, and a filterbank with low spectral resolution
which covered all $K$ components for the (6--5) and (8--7) transitions and up 
to the $K=10$ for the (13--12) line. In the following we shall make use
only of the spectra obtained with the autocorrelator because all lines
detected fall in the autocorrelator band.

For the \HCOpI\ (1--0) and the SiO (2--1), (3--2) and (5--4) ($v=0$)
lines, 3$\times$3 point maps were obtained, with
12$^{\prime\prime}$ (\HCOpI ) and 24$^{\prime\prime}$ (SiO) spacing.

We observed in ``wobbler switching'' mode, which uses a nutating secondary
reflector with a beam-throw of 240\asec\ in azimuth and a phase duration of
2 sec.
Pointing and receiver alignment were checked on nearby point sources, and 
were found to be
accurate within 2\asec . The data were calibrated with the
``chopper wheel'' technique. Calibration was checked by measuring the 
intensity of Venus at the same frequencies.
The main beam brightness temperature, \Tmb , and the flux density, 
$F_{\nu}$,
are related by the expression $F_{\nu}({\rm Jy})=4.9\;T_{\rm MB}({\rm K})$.
\subsubsection{Data reduction and fitting procedure}

\ace\ is a symmetric-top molecule whose rotational levels are described
by two quantum numbers: $J$, associated with the total angular momentum,
and $K$, associated with its projection on the symmetry axis. This structure
implies that for each $J+1\rightarrow J$ rotational transition, $J+1$ lines
are observable (for a detailed description see Townes \& Schawlow
\cite{townes}). To analyze the \ace\ data we have
adopted the fitting procedure described in Fontani et al.~(\cite{fonta}),
which assumes the
same rest velocity and linewidth for all $K$ components. 

All data were reduced with the GAG software developed 
at IRAM and Observatoire de Grenoble.

\subsection{NRAO observations}

Observations of the \CO\ (2--1) transition were made on February 21 1998 
towards \MI , and on January 10 1998 towards \MII , using the 12-m telescope
of the National Radio Astronomy Observatory (NRAO) at Kitt Peak.
We used the 1~mm SIS array receiver (Payne \& Jewell \cite{payne}), consisting
of $2\times 4$ independent beams. The maps were acquired with a 29\asec\ 
spatial
sampling, which is approximately the telescope HPBW at the \CO\ (2--1) line
frequency. The typical integration time on each position was about 4
minutes, with a system temperature of $\sim 1000$ K for \MI , and $\sim 2000$ 
K for \MII . Data were taken in position switching mode, and the same
reference positions given in Sect.~\ref{30m} were used, which are
CO-free down to $\sim 0.2$ K.
HPBW, spectral resolution and bandwidth are listed in Table~\ref{tobs}.

\subsection{JCMT observations}

The 850$\mu$m continuum images were taken on October 16 1998 with SCUBA at the
JCMT (Holland et al. \cite{holland}) towards the positions given in 
Table~\ref{tsources}. The standard 
64-points jiggle map observing mode was used, with a chop
throw of 2 arcmin in the SE direction. 
This results in a $3^{\prime}\times 3^{\prime}$ map size. The atmospheric
optical depth at 225 GHz was $\tau_{\rm 225 GHz} \sim$ 0.15. Telescope 
focus and pointing
were checked using Uranus and the data were calibrated following standard 
recipes as in the SCUBA User Manual (SURF).

\subsection{OVRO Observations}

Millimeter interferometric observations were performed using the
Owens Valley Radio Observatory millimeter array. \MI\ and \MII\
were observed in the period from October to December 2001.
The six 10.4-m dishes were employed in three separate configurations
(C, L, and H), offering baselines from the shadowing limit to
approximately 220~m. The flexible digital correlator was configured
to observe simultaneously the CO(1--0) and the C$^{17}$O(1--0) lines
at 2.6~mm. The resolution used was $\sim$0.9~\kms\ over a 80~\kms\
bandwidth for the CO(1--0), and $\sim$0.65~\kms\ over a 20~\kms\
bandwidth for the C$^{17}$O(1--0). Continuum observations have
been performed making use of the 2~GHz bandwidth of the analog
correlator. The observing cycle alternated observations of the
two target sources and of the phase calibrators (2201$+$508 and BL~Lac);
a calibration scan was acquired every $\sim$12~min. Pointing
was checked and refined by frequent pointing sessions on BL~Lac.
Bandpass calibration was performed using observations of 3C273 and/or
3C345, while the flux density scale was determined using observations
of either Uranus or Neptune (the expected accuracy is within 15\%).
The raw visibility data were calibrated using the OVRO in-house
MMA package (Scoville et al.~\cite{mma}); the calibrated source data
were then exported in FITS format for further analysis with standard packages.

\subsection{Near-Infrared observations}

\subsubsection{Palomar 60-inch broad-band observations}
\MI\ and \MII\ were observed in the three standard J, H, and K$_{\rm s}$
near-infrared broad-band filters using the P60IRC (Murphy et al.~\cite{Mea95})
camera at the
Palomar 60-inch telescope on 1999, November 30. For each source we
obtained a set of dithered frames. After flat fielding and sky subtraction
the frames were registered and coadded, producing final mosaics
of a region of $\sim$3\pam 5$\times$3\pam 5 around each
of the two IRAS sources. Photometric calibration was ensured by
observations of a set of stars from the Arnica and Las~Campanas
lists (Hunt et al.~\cite{Hea98}; Persson et al.~\cite{Pea98}).
The limiting magnitudes of our observations were found to be
18.0, 17.5 and 17.0 in J, H, and K$_{\rm s}$ respectively.
Astrometric calibration was performed using stars from the
HST Guide Star Catalogue, and later checked with 2MASS\footnote{http://www.ipac.caltech.edu/2mass/} objects in
the field (the derived accuracy is within 1$^{\prime\prime}$ rms).

\subsubsection{TNG/NICS narrow-band observations}

Near-Infrared narrow-band observations of \MI\ and \MII\ were performed on
August 21, 2002, at the 3.56-m Telescopio Nazionale Galileo (TNG) at the Roque
de Los Muchachos Observatory on the Spanish island of La Palma. The instrument
used is the Near Infrared Camera and Spectrograph (NICS, Baffa et
al.~\cite{Bea01}), a cryogenic focal reducer designed and built by the Arcetri
Observatory IR group as a common-user instrument for the TNG. NICS is equipped
with a Rockwell 1024$^2$ HAWAII near infrared array detector. The plate scale
used for our observations is 0\pas 252/pix, for a total field of
view of $\sim$4\pam 5. Series of dithered frames were obtained in the
FeII, H$_{\rm c}$, H$_2$, and K$_{\rm c}$ filters, centered at the rest 
frequencies of the
[FeII]~$\lambda\sim$1.64~$\mu$m transition, the H$_2$(1-0)S(1)
$\lambda\sim$2.12~$\mu$m transition and the nearby continua (see Ghinassi et
al.~\cite{Gea02} for more details on the NICS filters). The seeing at the time
of the observations was $\sim$0\pas 9, but the sky conditions were
not photometric, which did not allow to obtain a flux calibration for our
images.

Data reduction and analysis were performed using the IRAF software package,
following standard flat-fielding and sky subtraction, the individual images
were registered and the final mosaic was produced. The narrow-band
``continuum'' filters were used to subtract continuum emission from the FeII
and H$_2$ images, assuming that a set of stars within each frame were not
affected by significant line emission or absorption. Note that, because the
H$_{\rm c}$ filter has a central wavelength slightly shorter ($\sim$1.57~$\mu$m) than
that of the [FeII] filter, while K$_{\rm c}$ has a central wavelength slightly longer
($\sim$2.19~$\mu$m) than the H$_2$ filter, very red stars will leave a positive
residual in the [FeII] continuum subtracted image and a negative residual in
the H$_2$ continuum subtracted image. Accurate ($\le 0\farcs 5$) astrometry
was derived for all mosaics using stellar positions from the 2MASS
observations.

\section{Results}
\label{res}
\begin{table*}
\begin{center}
\caption[] {Observed sources: the distance, $d$, and the bolometric 
luminosity, $L$, are taken from Molinari et al. (\cite{mol02}).
The source velocity, $v_{\rm LSR}$, corresponds to the molecular gas velocity,
 derived from \AMM\ emission (Molinari et al. \cite{mol96}).} 
\label{tsources}
\begin{tabular}{c|cccccc}
\hline
Name  & R.A.(J2000) & Dec.(J2000) & $d$ & $R_{\rm GC}$ & $L$ & $v_{\rm LSR}$ \\
      &             &             & (kpc)& (kpc) & ($\times 10^{3} L_{\odot}$) & (\kms )\\
\hline
I21307 & 21$^{h}32^{m}30.6^{s}$ & 51$^{\circ}02^{\prime}16.5^{\prime\prime}$ &
3.2 & 8.4 & 4.0 & $-$46.7  \\
I22127 & 22$^{h}19^{m}08.6^{s}$ & 56$^{\circ}05^{\prime}02.0^{\prime\prime}$ &
2.4 & 7.8 & 1.8 & $-$43.8 \\
\hline
\end{tabular}
\end{center}
\end{table*}

\subsection{Line emission}
\label{spec}

\begin{table*}
\begin{center}
\caption[] {Observed molecular transitions.}
\label{tobs}
\begin{tabular}{c|c|c|c|c|c}
\hline \hline
 Molecular   &  Rest freq. & Telescope   & HPBW & Bandwidth & Resolution \\
  transition    &    (GHz) &   & (\asec )  & (MHz) & (MHz)  \\
\hline
\CO\ (1--0) & 115.271  & OVRO & 6.95$\times$6.12 & 26.3 & 0.33 \\
\CO\ (2--1) & 230.538  & 30m  & 11  & 320  & 0.32 \\
\CO\ (2--1)  & 230.538    & NRAO & 29 & 150 & 0.78 \\
\CII\ (2--1) & 219.560   &  30m & 12 & 140 & 0.078 \\
\HCOpI\ (1--0) & 86.754 & 30m &  29  &  70 & 0.078 \\
\ace\ (6--5) & 102.547$(^*)$ & 30m &  24  &   70 & 0.078  \\ 
\ace\ (8--7) & 136.728$(^*)$ &  30m &  18  &   70 & 0.078  \\
\ace\ (13--12) & 222.166$(^*)$ &  30m &  11  &   140 & 0.078 \\
SiO ($v=0$,$J$=2--1) & 86.846 & 30m & 29 & 70 & 0.078 \\
SiO ($v=0$,$J$=3--2) & 130.269 & 30m & 20 & 70 & 0.078 \\
SiO ($v=0$,$J$=5--4) & 217.105 & 30m & 12 & 140 & 0.078 \\  
\hline
\end{tabular}
\end{center}
$(^*)$ rest frequency of the $K=0$ transition
\end{table*}

\begin{figure}
\centerline{\includegraphics[angle=0,width=7.5cm]{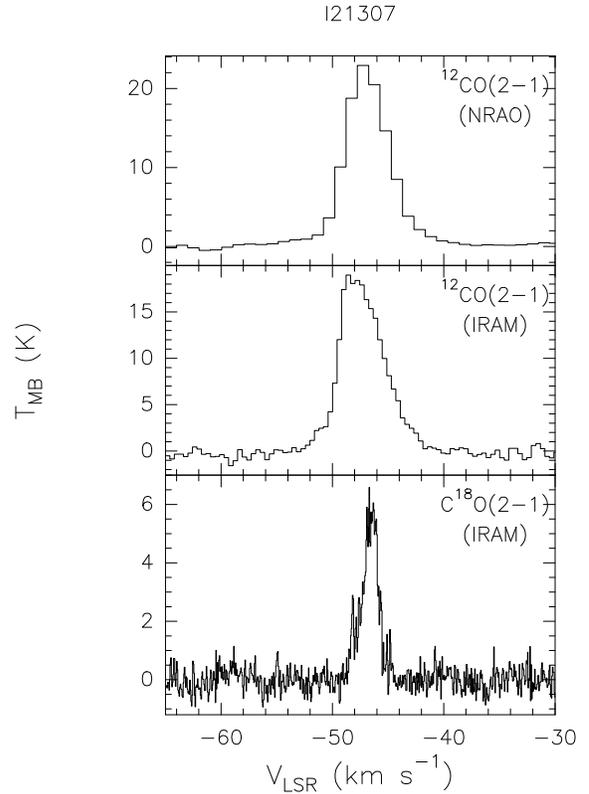}}
\caption{From top to bottom: spectra of the \CO\ (2--1) line towards \MI , 
observed with the 
NRAO-12m and the IRAM-30m telescopes, and of the \CII\ (2--1) line
observed with the IRAM-30m telescope. All have been taken at the peak 
position in the maps. }
\label{spec_mol136}
\end{figure}

\begin{figure}
\centerline{\includegraphics[angle=0,width=7.5cm]{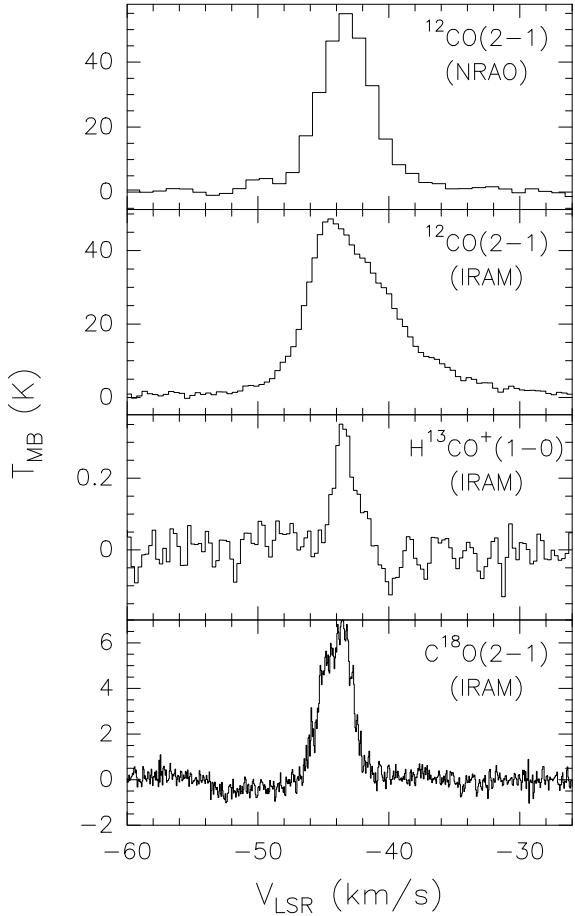}}
\caption{Spectra of the \CO\ (2--1) line observed with the IRAM-30m telescope
and the NRAO telescope, and of the 
\HCOpI\ (1--0) and \CII\ (2--1) lines observed with the IRAM-30m telescope. 
All spectra refer to the peak position in the maps.}
\label{spec_mol143}
\end{figure}

\begin{figure}
\centerline{\includegraphics[angle=0,width=7.5cm]{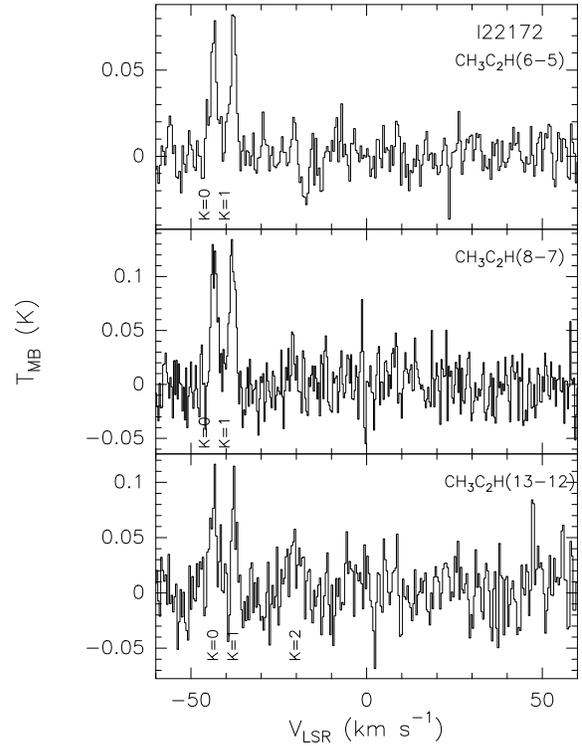}}
\caption{From top to bottom: spectra of the \ace\ (6--5), (8--7) and (13--12) 
lines observed towards \MI\ with the IRAM-30m telescope. The numbers under
the spectra indicate the position of the different $K$ components. The 
$V_{\rm LSR}$ is relative to the $K=0$ line.}
\label{spec_ace}
\end{figure}

In Figs.~\ref{spec_mol136} and \ref{spec_mol143}, we show the \CO\ (2--1),
\CII\ (2--1) and \HCOpI\ (1--0) (not observed towards \MI ) spectra taken at 
the central position in the maps. We also show the 
\ace\ (6--5), (8--7) and (13--12) spectra obtained towards \MII\ 
(Fig.~\ref{spec_ace}): for these lines, only single-pointing spectra are
available. 
The \CO\ (2--1) lines show non-Gaussian wings, while the \CII\ (2--1) line,
the \HCOpI\ (1--0) line and the \ace\ lines are well fitted by Gaussians.
The SiO (2--1), (3--2) and (5--4) ($v=0$) lines are 
undetected up to a $3\sigma$ level of 0.18 K, 0.4 K and 0.3 K, respectively.

\begin{figure}
\centerline{\includegraphics[angle=0,width=7.5cm]{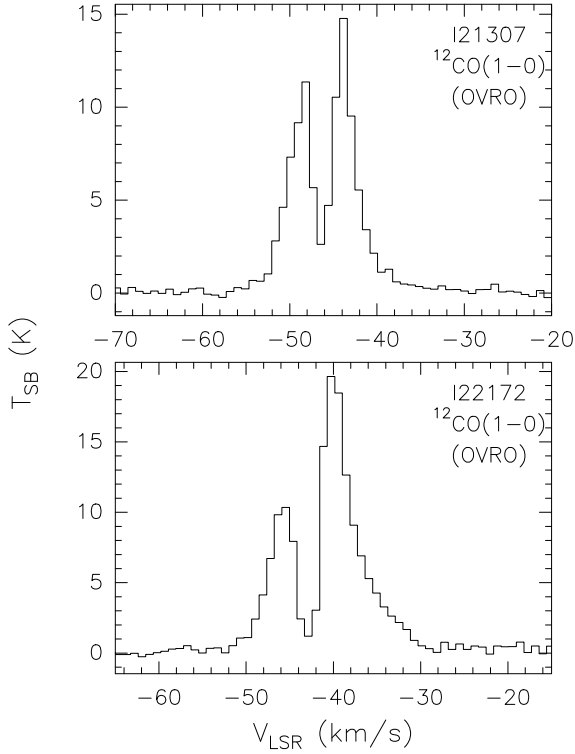}}
\caption{Top panel: spectrum of the \CO\ (1--0) line observed with the OVRO
interferometer towards \MI . The spectrum has been obtained integrating the 
emission over the 3$\sigma$ level. Bottom panel: same as top panel for \MII . }
\label{spec_ovro}
\end{figure}

Finally, Fig.~\ref{spec_ovro} presents the spectra of the
\CO\ (1--0) line observed with the OVRO interferometer: both spectra have
been obtained by integrating the emission of the line above the 
$3\sigma$ level. In this case, the brightness temperature in the 
synthesized beam,
$T_{\rm SB}$, is plotted against the velocity. Both spectra clearly show 
intense emission in the wings, and the central peak is lacking: this is 
due to
the fact that the channels near the line peak are representative of the 
more extended emission, which has been resolved out by the 
interferometer. 

\begin{figure}
\centerline{\includegraphics[angle=0,width=7.5cm]{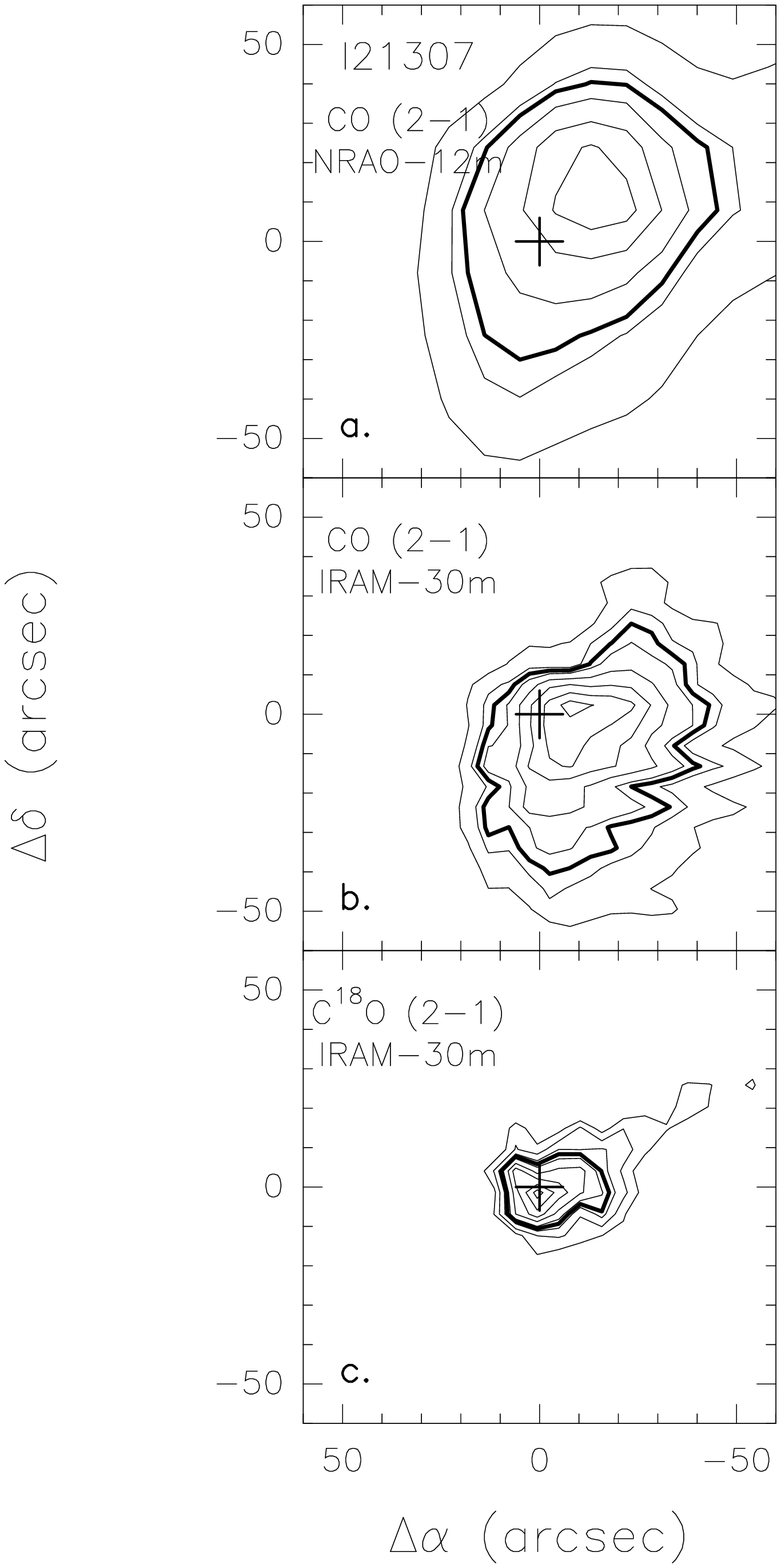}}
\caption{{\bf a}: Map of the \CO\ (2--1) line integrated 
between $-55$ and $-38$ \kms , observed towards \MI\ obtained with the NRAO-12m telescope. Contour levels range from
60 to 200 by 30 K \kms . The thick contour indicates the half maximum power
contour. The cross at the map center shows the position of the sub-mm peak
(Molinari et al. \cite{mol00}).
{\bf b}: Same as {\bf a} for the IRAM-30m telescope. Same contour 
levels are shown. {\bf c}: Same as {\bf a} for the \CII\ (2--1) line 
integrated between $-50$ and $-42$ \kms , observed with
the IRAM-30m telescope. Levels range from 2.2 to 9.2 by 1.0
\kms .}
\label{comap_m136}
\end{figure}

\begin{figure}
\centerline{\includegraphics[angle=0,width=7.5cm]{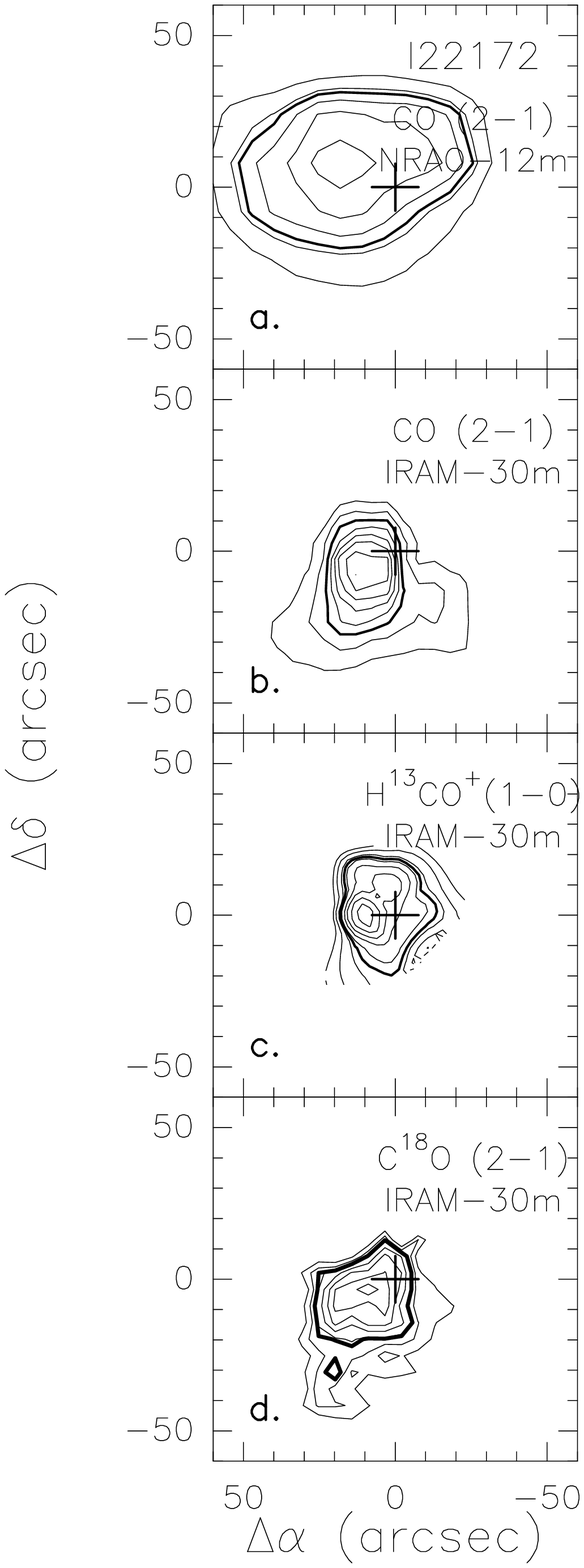}}
\caption{{\bf a}: Map of the \CO\ (2--1) line integrated between $-51$ and
$-30$ \kms , observed towards \MII\ with the NRAO-12m telescope. Contour levels
range from 80 to 255 by 35 K \kms . The thick contour indicates the half 
maximum power
level. The cross at the map center shows the position of the 3.4~mm peak
(Molinari et al. \cite{mol02}). {\bf b}: Same as {\bf a} for the 
IRAM-30m telescope. The same contour levels are shown. {\bf c}: Same as 
{\bf a} for the \HCOpI (1--0) line,
observed with the IRAM-30m telescope, integrated between $-46$ and $-40$ \kms .
 Contour levels range from 0.18 to 0.74
by 0.07 K \kms .{\bf d}: Same as {\bf c} for the \CII\ (2--1) line.
Levels range from 2.4 to 9.0 by 1.1 \kms .} 
\label{comap_m143}
\end{figure}

The integrated maps of the \CO\ (2--1), \CII\ (2--1) and \HCOpI\ (1--0) lines
are presented in Figs.~\ref{comap_m136} and \ref{comap_m143}. 
In \MI, the \CII\ integrated intensity map peaks right
on the sub-mm source detected by Molinari et al.~(\cite{mol00}). In \MII\ 
all line maps are offset by $\sim10$\asec\ 
E from the millimeter core (Molinari et al. \cite{mol02}).
Since the \HCOpI\ and \CII\ observations have been carried out
in different observing runs, it is unlikely that this effect is due to
a pointing error. Moreover, we will see in Sect.~\ref{morf}
that a similar offset between line and continuum
is confirmed by the interferometric observations.

In Table~\ref{tline} we give the outflow parameters, namely: 
 the velocity range of the blue- and red- wings, 
$\Delta v_{\rm b}$ and $\Delta v_{\rm r}$ (Cols.~3 and 4, respectively), the 
integrated intensity in the wings, $\int T{\rm d}v$ (blue) and (red)
(Cols.~5 and 6), and the maximum velocity of the wings, $v_{\rm max_{b}}$ and 
$v_{\rm max_{r}}$ (Cols.~7 and 8), defined as the difference
between the maximum velocity of the blue- and red- wings and the line peak.
The latter has been derived from the rest velocity given in 
Table~\ref{tsources}. 
The velocity ranges of the wings have been determined from the spectrum 
of the \CO\ (1--0) line observed with OVRO, which has the best angular 
resolution, and they are consistent with the wings observed in the other
lines. The values listed in Cols.~5 and 6
are mean brightness temperatures over the blue and the red lobes, respectively.

\begin{table*}
\begin{center}
\caption[] {Line parameters for outflow calculations. }
\label{tline}
\begin{tabular}{cccccccc}
\hline \hline
Source & Line & $\Delta v_{\rm b}$ & $\Delta v_{\rm r}$ & 
$\int T{\rm d}v$ (blue) & $\int T{\rm d}v$ (red) & $v_{\rm max_{b}}$ & $v_{\rm max_{r}}$ \\
  &  & (\kms\ )  &  (\kms\ ) & (K \kms\ ) & (K \kms\ )  & (\kms\ )  &(\kms\ )   \\
\hline
\MI\  & \CO\ (2--1) (NRAO) & ($-55$, $-50$)  & ($-44$, $-38$) & 10.2 & 8.9 & 8.3 & 8.7 \\
      & \CO\ (2--1) (30m) & ($-55$, $-50$)   & ($-44$, $-38$) & 16.2 & 14.9 & 8.3 & 8.7  \\
      &\CO\ (1--0) (OVRO) &  ($-55$, $-50$)  &  ($-44$, $-38$) & 13.2 & 34.8 &  8.3  & 8.7 \\
\MII\  & \CO\ (2--1) (NRAO) & ($-51$, $-45$)  & ($-38$, $-30 $) & 21.1 & 22.5 & 7.4 & 13.6 \\
      & \CO\ (2--1) (30m) & ($-51$, $-45$)  & ($-38$, $-30$) & 25.4 & 30.0 & 7.4 & 13.6 \\
      &\CO\ (1--0) (OVRO) & ($-51$, $-45$)  &  ($-38$, $-30$) & 37.2 & 38.1 &  7.4 &  13.6 \\
\hline
\end{tabular}
\end{center}
\end{table*}


\subsection{Continuum maps}
\label{cont}

\begin{figure}
\centerline{\includegraphics[angle=-90,width=7.5cm]{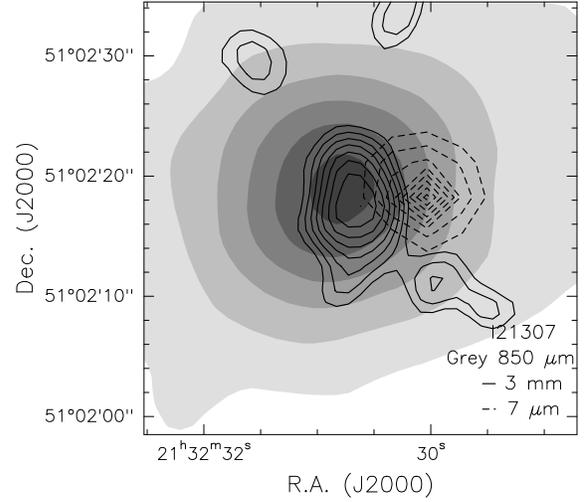}}
\caption{SCUBA map of \MI\ at 850 $\mu$m 
(grey-scale) superimposed to the 3~mm OVRO map (solid lines) and the 
7 $\mu$m ISOCAM map (dashed lines). Contour levels range from: 
0.2 ($4\sigma$) to 1.1 by 0.16 Jy beam$^{-1}$ in the SCUBA map; from 0.003 
($\sim3\sigma$) to 0.015 by 0.001 Jy beam$^{-1}$ in the OVRO map; from 20
to 212 by 24 Jy beam$^{-1}$ in the ISOCAM map.}
\label{m136_map_tot_isocam}
\end{figure}

\begin{figure}
\centerline{\includegraphics[angle=0,width=7.5cm]{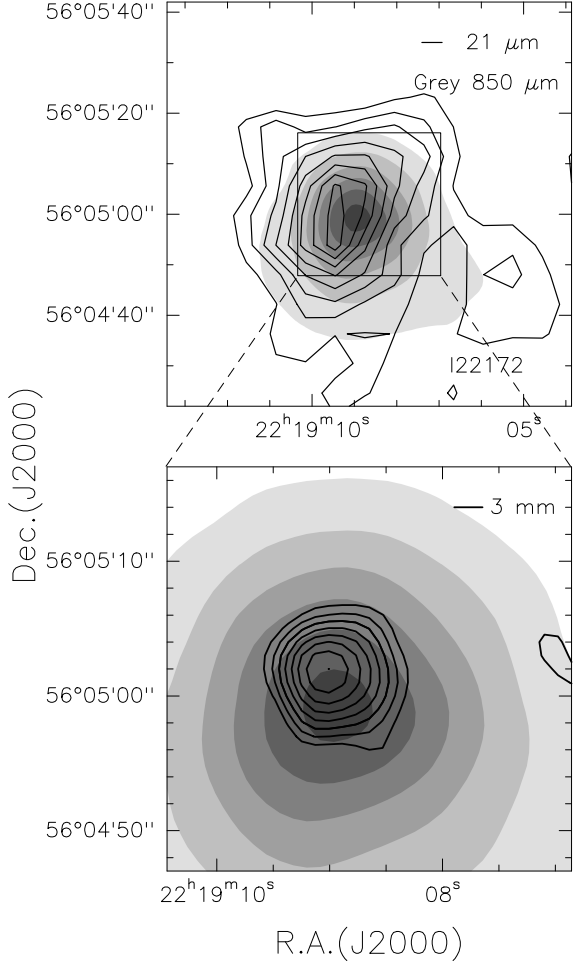}}
\caption{Top panel: 850 $\mu$m SCUBA image of \MII\ (grey scale)
superimposed to the 21 $\mu$m MSX map (solid lines). Contour levels
of the SCUBA map range from 0.3 ($3\sigma$) to 1.5 by 0.24 Jy beam$^{-1}$.
Bottom panel: enlargement of the SCUBA map around the central 
region. Contours indicate the OVRO 3~mm emission, whose
levels range from 0.004 ($\sim3\sigma$)
to 0.016 by 0.001 Jy beam$^{-1}$.}
\label{m143_map_tot}
\end{figure}

\begin{table*}
\begin{center}
\caption[] {Physical parameters of the \MI\ core. $\theta_{\rm s}$ and $D$ 
are the angular and linear diameter, respectively, computed for a distance of
3.2 kpc.}
\label{tm136_core}
\begin{tabular}{c|ccccc}
\hline
   & $\theta_{\rm s}$  & D & $M_{\rm vir}$ & $M_{\rm H_2}$ & $n_{\rm H_{2}}$ \\
   & (arcsec)   & (pc) & ($M_\odot$) &  ($M_\odot$) & (cm$^{-3}$ ) \\
\hline
\CO\ (2--1) (12m)  & 63 & 0.97 & 2061  & -- & -- \\
\CO\ (2--1) (30m)  & 40 & 0.62 & 1640  & --  & --  \\
\CII\ (2--1) (30m) & 18 & 0.28  & 360  & 62  & $4.0\times 10^{5}$  \\
\CO\ (1--0) (OVRO) & 14 & 0.24 & -- & -- & -- \\
850 $\mu$m (SCUBA) &  8.6  & 0.13 &  -- &  155 & $9.9\times 10^{6}$  \\
3~mm (OVRO)  & 5.8   &  0.09   & --  & 53 & $9.3\times 10^6$ \\
\hline
\end{tabular}
\end{center}
\end{table*}

\begin{table*}
\begin{center}
\caption[] {Physical parameters of the \MII\ core. $\theta_{\rm s}$ and $D$
are computed for a distance of 2.4 kpc.}
\label{tm143_core}
\begin{tabular}{c|ccccc}
\hline
       &   $\theta_{\rm s}$ & D & $M_{\rm vir}$ & $M_{\rm H_2}$ & $n_{\rm H_{2}}$  \\
      &   (\asec ) & (pc) & ($M_\odot$) & ($M_\odot$) & (cm$^{-3}$) \\
\hline
\CO\ (2--1) (12m) & 51 & 0.59 & 1330 & -- & -- \\
\CO\ (2--1) (30m) & 30 & 0.35 & 780 & -- & -- \\
\CII\ (2--1) (30m) & 25 & 0.29  & 250  & 95  & $5.1\times 10^{5}$ \\
${\rm H^{13}CO^{+}}$ (1--0) (30m) & 15 & 0.17 & 68 & 120 & $3.1\times 10^{6}$ \\
\CO\ (1--0) (OVRO) & 14 & 0.16 & -- & -- & -- \\
850 $\mu$m (SCUBA) & 11 & 0.13 & -- & 128 & 8.2 $10^{6}$ \\
3~mm (OVRO) & 3.34 & 0.04 & -- & 38 & 7.6 $10^{7}$ \\
\hline
\end{tabular}
\end{center}
\end{table*}
In Fig.~\ref{m136_map_tot_isocam}, the 850 $\mu$m map observed with SCUBA
towards \MI\ is superimposed on the 3~mm and the 7 $\mu$m emission 
observed with OVRO and ISOCAM, 
respectively. The angular resolution of both the ISOCAM and OVRO maps is 
$\sim6$\asec , identical to that of the 
\CO\ (1--0) observations (Table~\ref{tobs}).
The intensity profile of the SCUBA map indicates the presence of an
unresolved central core, and a halo extending over $\simeq 40$\asec .
Such a halo is not detected in the line maps presented in Sect.~\ref{spec}.
However, it is well known that the optically thin emission of dust and 
the molecular line intensities do not depend on temperature and column 
density in the same manner. Furthermore, chemistry and optical depth
effects can affect the line emission and cause a different shape of the maps.
  
The OVRO map indicates the presence of a compact
core, whose diameter is $\sim 0.09$ pc, located near the emission peak of the 
850 $\mu$m map. One sees that the emission peak of
the ISOCAM map is offset by $\sim6$\asec\ in R.A. from the the 3~mm and
the 850 $\mu$m emission peaks. This is confirmed by the
15 $\mu$m ISOCAM map, whose emission peak is nearly coincident with that
detected at 7 $\mu$m.
Therefore, the ISOCAM maps may suggest that in \MI\ the bulk of the MIR 
emission does not arise from the millimeter core. However, we point out
that these maps have to be regarded with caution because the 
astrometric uncertainty may be as large as $\sim10$\asec\ 
(see Cesarsky \& Blommaert \cite{cesarsky}), and maps with
better astrometric precision and angular resolution are required to
assess the origin of the mid-infrared emission. 

The top panel of Fig.~\ref{m143_map_tot} shows the SCUBA map observed towards
\MII . In this case, to make a comparison with the MIR emission, we have 
superimposed the MSX\footnote{ MSX images have been taken from the
on-line MSX database http://www.ipac.caltech.edu/ipac/msx/msx.html}
21 $\mu$m image, because no ISOCAM maps are available for this source. The 
bottom panel presents an enlargement of the central
region with a contour plot of the 3~mm continuum observed with OVRO,
which shows the presence of a compact millimeter core ($\sim 0.04$ pc)
approximately located at the center of the SCUBA map.
Although the angular resolution of the MSX map is $\sim4$ times worse
than that of ISOCAM, we find that also in this case the peak of 
the MIR emission does not coincide with that at 850 $\mu$m and
3~mm.

\begin{figure}
\centerline{\includegraphics[angle=0,width=7.5cm]{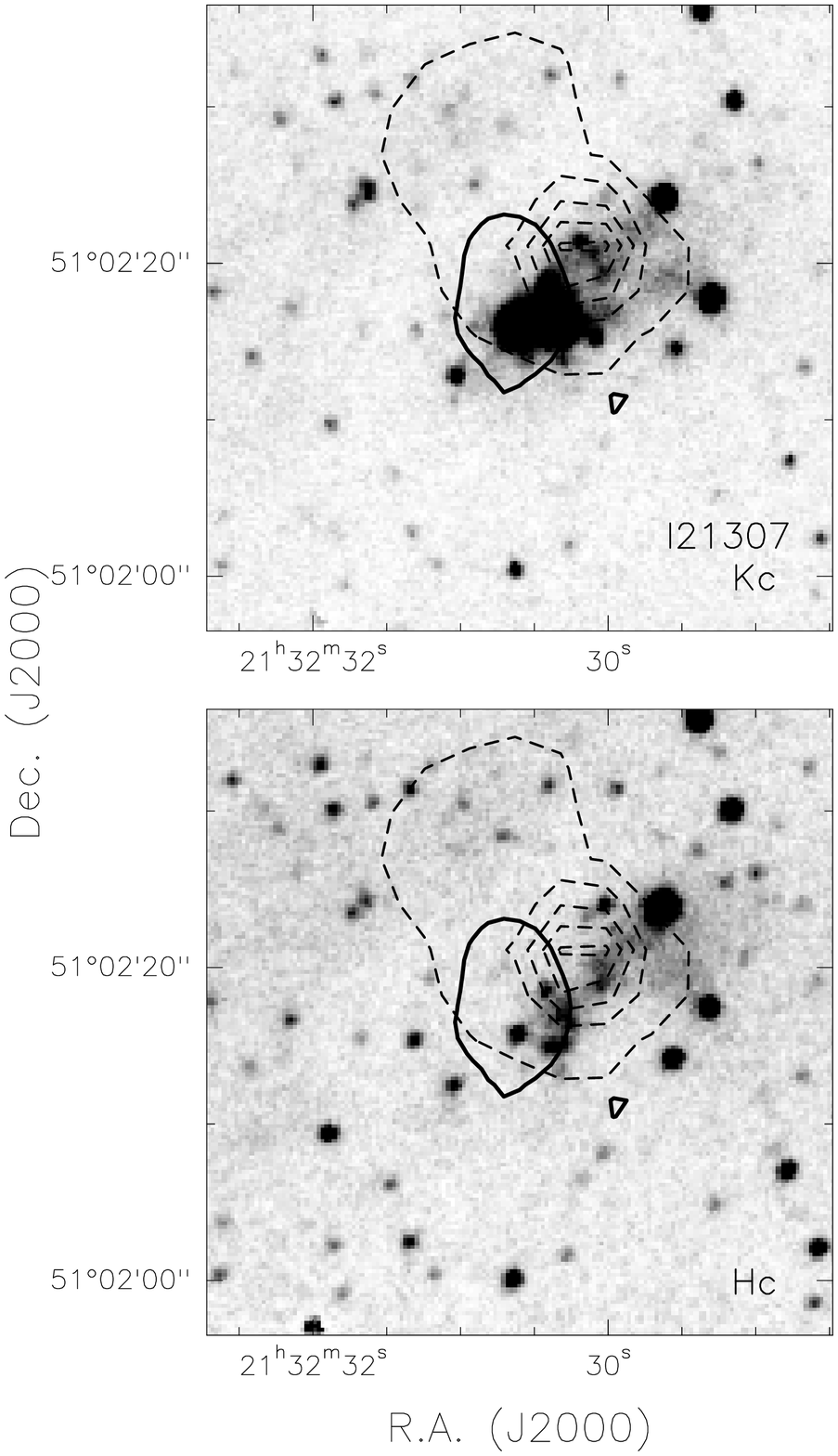}}
\caption{
Kc-band (top panel) and Hc-band (bottom panel) images of \MI . 
The position of the 3~mm core is
indicated by the solid line, which represents the
half maximum power contour of the OVRO maps shown in 
Fig.~\ref{m136_map_tot_isocam}. The dashed lines indicate the 
ISOCAM 15 $\mu$m image. The same contour levels as in 
Fig.~\ref{m136_map_tot_isocam} are shown.}
\label{m136_nir}
\end{figure}

\begin{figure}
\centerline{\includegraphics[angle=0,width=7.5cm]{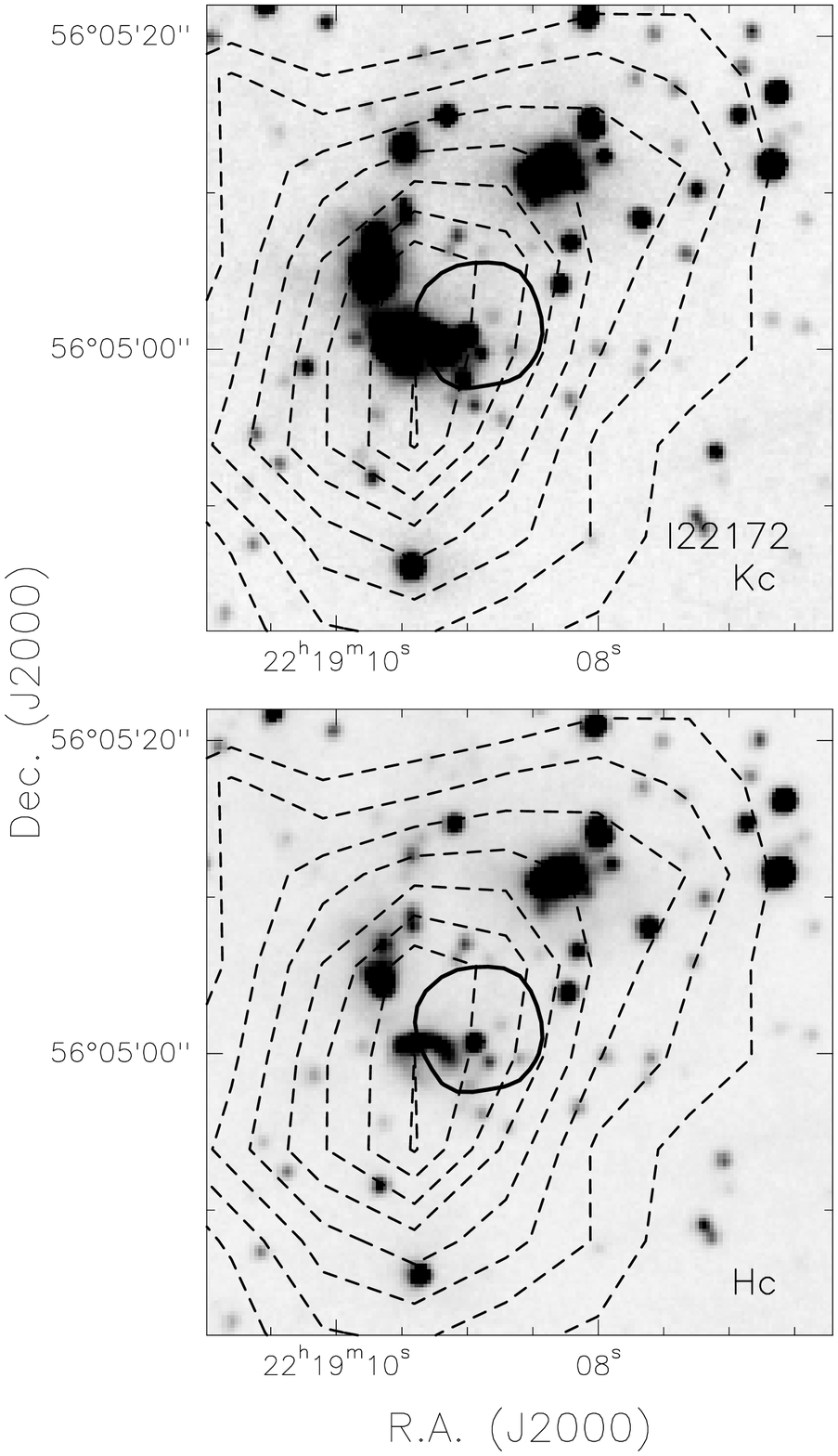}}
\caption{Same as Fig.~\ref{m136_nir} for \MII . The dashed lines represent 
the MSX 21 $\mu$m image, with the same contour levels as in 
Fig.~\ref{m143_map_tot}.}
\label{m143_nir}
\end{figure}

Near-infrared images of both sources are presented in Fig.~\ref{m136_nir}  
and \ref{m143_nir}, obtained at 2.2 $\mu$m (Kc-band) and 1.7 $\mu$m (Hc-band).
In both maps, the half maximum power 
contour of the 3~mm core, and the MIR emission are also shown. The Kc-band 
image of \MI\ (top panel of Fig.~\ref{m136_nir}) shows a cluster of stars,
the ``centroid'' of which is located at 
$\sim6$\asec\ ($\sim 0.08$ pc) in R.A. from the peak of the millimeter 
core. However, the brightest cluster member in K$_{\rm c}$ is $\sim 1$\asec\
South of the OVRO peak. The 15 $\mu$m ISOCAM peak lies on the western 
side of the cluster. 
It seems probable to us that such emission is due to a
combination of emission from the cluster members and small particle
emission from the adjacent cloud.

In \MII , the 3~mm core is nearly at the center of a
cluster of NIR sources distributed as a patchy ``ring'', detected both 
at 1.7 and 2.2 $\mu$m (Fig.~\ref{m143_nir}). In this case also the 
21 $\mu$m emission is roughly coincident with the position of the cluster. 

It is interesting to notice that in both sources the stars
detected in the near-infrared images are located close to the millimeter
core, but have a distribution that does not
exactly coincide with it: the ``average position'' of the reddest stars
of the cluster is $\sim6-8$\asec\ offset from the millimeter peak, i.e.
well above the positional uncertainties on the maps, estimated to be
$< 1$\asec . This is reminescent of the observations of
the candidate massive protostar IRAS 23385+6053 (Fontani et al. 
\cite{fonta2}), in which a compact massive core, believed to host a newly
formed massive (proto)star, is surrounded by a cluster
of more evolved B stars, distributed as a patchy ring around the
core. 

\subsubsection{The NIR clusters}
\label{srclus}

The near infrared broad band images of \MI\ and \MII\ clearly show a 
concentration of red star-like sources near the IRAS positions.
This is consistent with the detection of young stellar objects 
closely associated with the molecular clumps. Following the methods
described in Testi et al.~(\cite{Tea97}; \cite{Tea98}) we have verified
the presence of a stellar surface density peak and we have estimated 
the {\it richness indicator} I$_{\rm C}$ for both regions.
The resulting values are I$_{\rm C}\sim$13$\pm 3$ for \MI\ and 
I$_{\rm C}\sim$9$\pm 3$
for \MII . 

In order to compare these numbers with other cluster surveys, it
is necessary to give a rough estimate of the minimum stellar mass we are
able to detect in our observations.
Following Testi et al.~(\cite{Tea98}), but using the Palla \&
Stahler~(\cite{PS99}) pre-main sequence evolutionary tracks, we have computed
the expected stellar mass sensitivity of our observations. Neglecting 
the effects of infrared excess and assuming an age of $\sim$ 0.1--1~Myr
and an extinction A$_V$=10 mag, we estimate that our observations
should be sensitive to young stellar objects more massive than
$\sim$0.5--1.5~M$_\odot$.
This is necessarily a rough estimate, and our observations are certainly
not complete to these low masses. 

\subsection{SED and luminosities}
\label{lum}
\begin{figure}
\centerline{\includegraphics[angle=0,width=8cm]{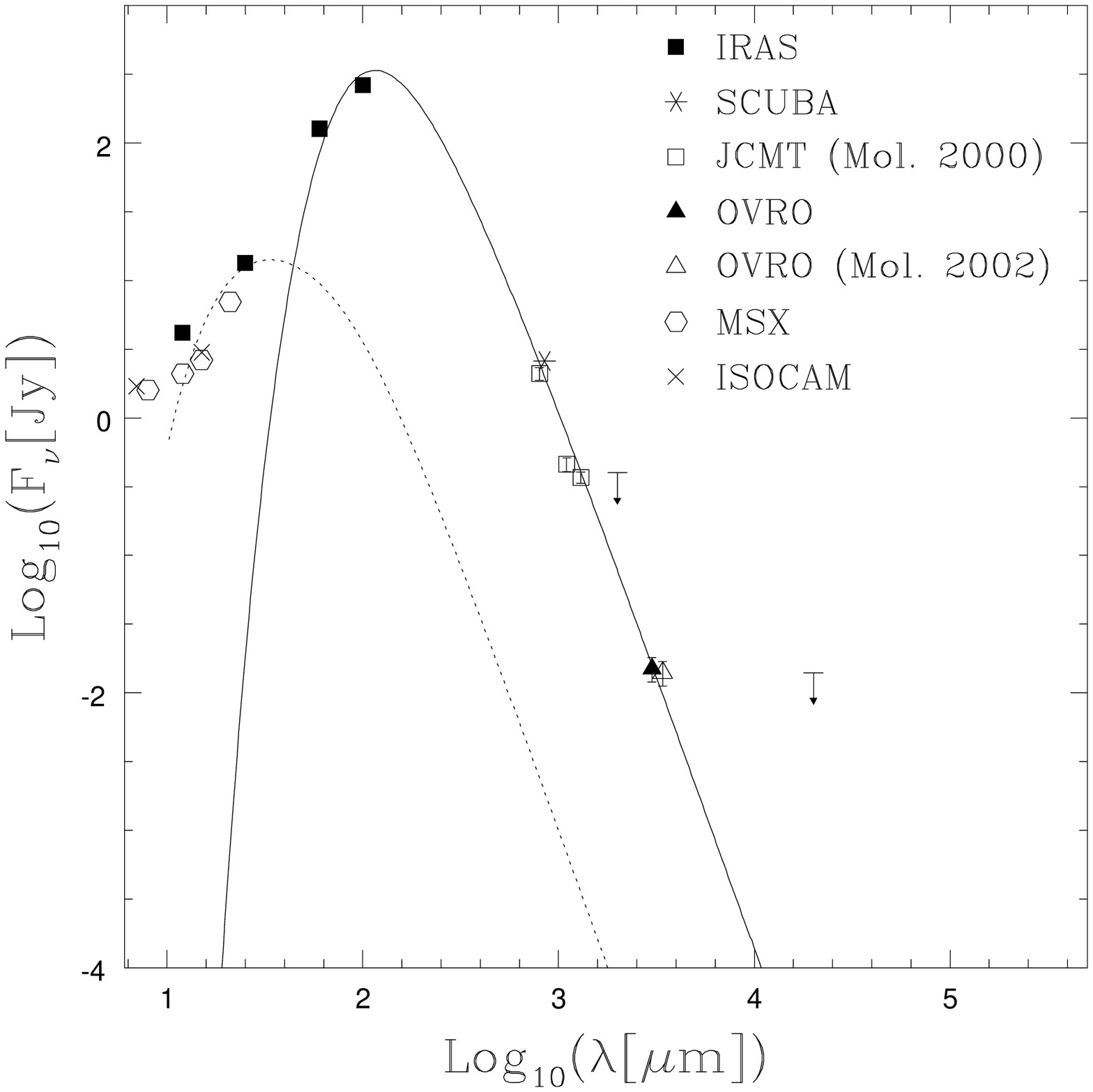}}
\caption{SED of \MI . The meaning of the symbols is given in the Figure. 
The arrows at 2~mm and 2~cm represent the upper limits ($3\sigma$ 
level) of the JCMT and VLA maps, respectively. The solid line represents a 
grey-body fit to the far-infrared and
millimeter points, with dust temperature
$T=30$ K and opacity index $\beta=2$. The dotted line is a grey-body fit 
to the data at $\lambda\leq25\;\mu$m, whit dust temperature $T=150$ K.}
\label{fsed_m136}
\end{figure}

\begin{figure}
\centerline{\includegraphics[angle=0,width=8cm]{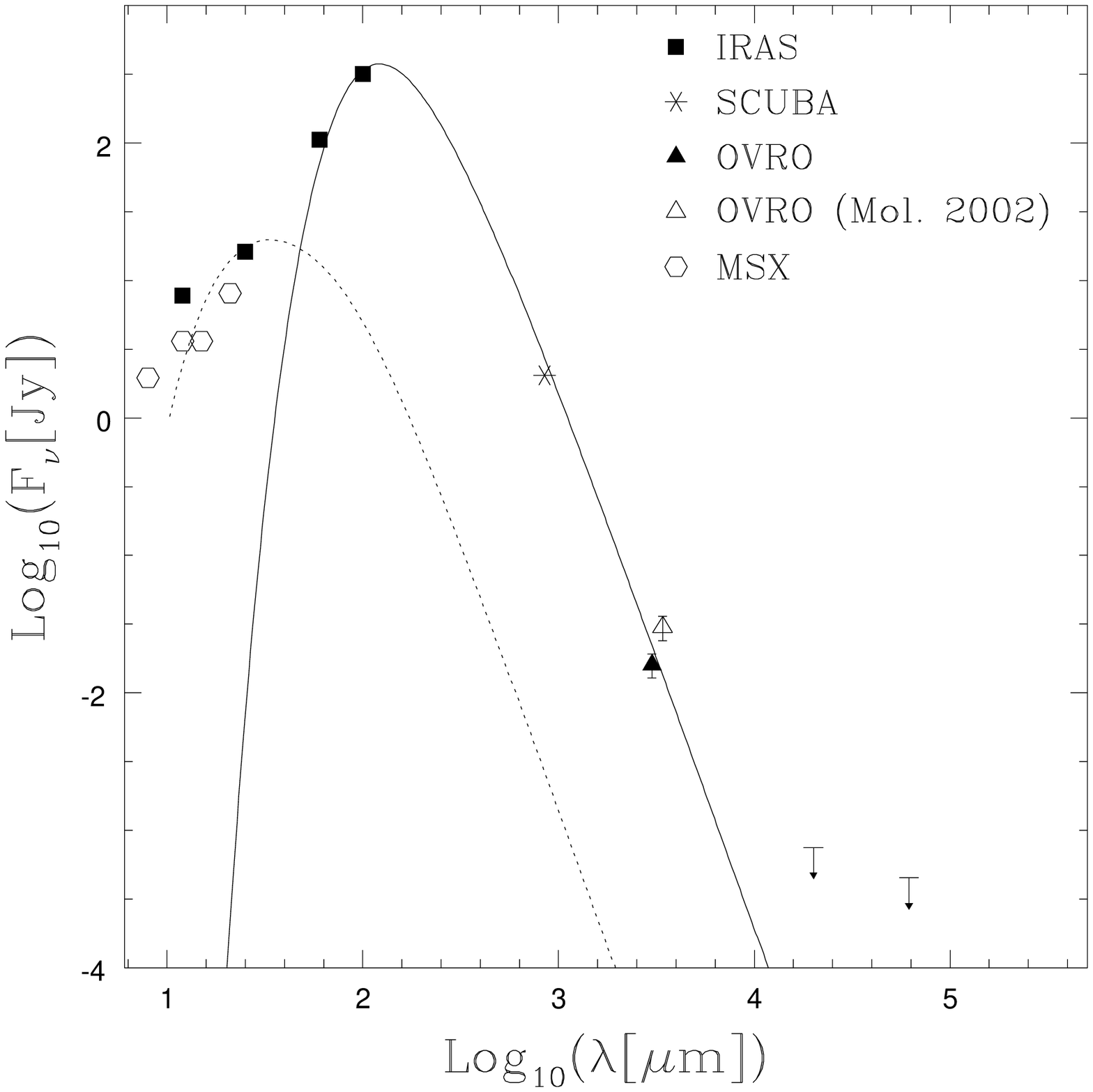}}
\caption{SED of \MII . Symbols have the same meaning as in
Fig.~\ref{fsed_m136}. The arrows at 2~cm and 6~cm represent the upper limits
($3\sigma$ level) of the VLA maps. The solid line is a grey-body fit to the 
far-infrared 
and millimeter points, which yields a dust temperature $T=27$ K, 
and an opacity index $\beta=2$. The dotted line is a grey-body fit 
to data with $\lambda\leq25\;\mu$m, with dust temperature $T=140$ K.} 
\label{fsed_m143}
\end{figure}

In Figs.~\ref{fsed_m136} and \ref{fsed_m143} we plot the Spectral Energy
Distribution (SED) of both sources.

It is clear that these regions are very complex and a proper
modeling of the SED would require a detailed radiation transfer 
calculation, taking into account both the geometry of the core and 
cluster as well as the radiation field of each cluster member
and the contribution of very small dust grains and PAHs.
This approach would require a substantial number of assumptions and
goes beyond the scope of this paper. Our
aim here is to estimate the relative contribution to the SED of the
two main components detected in each region: the NIR cluster and the
cold molecular core. This simplified approach is described in the
following.

Let us consider first \MI : the SED of this source was already 
derived by Molinari et al. (\cite{mol00}). With respect to 
their Fig.~3, Fig.~\ref{fsed_m136} contains more data, namely MSX, IRAS, 
OVRO and VLA. 
On the basis of the observations presented in 
Sect.~\ref{cont} we can distinguish between two components: 
the 3~mm core (hereafter I21307-C), and the
emission of a stellar cluster, detected in the NIR images. 
The millimeter and sub-millimeter flux
densities arise from the core I21307-C,
while, as discussed in Sect.~\ref{cont}, the MIR data are likely due to
the nearby cluster.  
Therefore in Fig.~\ref{fsed_m136} we have performed two grey-body fits.

The solid line represents the best grey-body fit obtained using data
representative of the emission arising from I21307-C, i.e. from millimeter to 
far-infrared wavelengths. 
Assuming a dust opacity index $\beta=2$, which is a value
typically found in envelopes of high-mass YSOs 
(Hunter 1997; Molinari et al.~\cite{mol00})
the corresponding parameters are: dust temperature $T=30$ K, angular
diameter $\theta=10$\asec , and mass $M\sim 90$ $M_{\odot}$.
This fit yields a luminosity of $4\times 10^{3} L_{\odot}$, which corresponds
to the luminosity of a B1.5 ZAMS star. This luminosity
has to be regarded as an upper limit to the ``real'' luminosity of the core
I21307-C. This is because the IRAS flux densities at 
100 and 60 $\mu$m may arise both from the 
core and from the external cluster, because the IRAS beam is much larger than 
the core region. However, the goodness of the grey-body fit shown in 
Fig.~\ref{fsed_m136} suggests that the 100 and 60 $\mu$m flux densities
are mainly due to I21307-C, and therefore the luminosity computed above
is close to the ``real'' core luminosity.

The dotted line is a grey-body fit to the data at wavelengths 
$\lambda\leq25\;\mu$m, and it represents the emission coming from
the external cluster. The fit is obtained for a temperature of 150 K.
By integrating the flux densities we obtain a 
luminosity of $3.6\times 10^{3} L_{\odot}$.

Let us now consider \MII : Fig.~\ref{fsed_m143} shows the SED of this source. 
The images obtained at different wavelengths show that also in this
case there is a molecular core (hereafter I22172-C), detected in 
the millimeter continuum and molecular lines maps, and a cluster of stars 
revealed in the near-infrared images. Therefore, as for I21307-C, we have 
performed a two grey-body fit to the SED.
Using $\beta=2$, from
Fig.~\ref{fsed_m143} we have derived the following best fit parameters: dust 
temperature $T=27$ K, angular diameter of the source $=9$\asec\ and
mass of $\sim 83$ $M_{\odot}$. The resulting luminosity is 
$2.2\times 10^{3} L_{\odot}$. In this case also, this luminosity must be
regarded as an upper limit for the luminosity of I22172-C, 
but, as discussed for \MI , the fact that the fit is reasonable suggests
that the emission at 100 and 60 $\mu$m arises
mainly from the core, so that the real core luminosity is close to the
value estimated before.
The dotted line shows the grey-body fit to data with $\lambda\leq25\;\mu$m, 
which represent the
emission due to the cluster of stars surrounding the core. It has been obtained
for a temperature of 140 K and a luminosity of $2.2 \times 10^{2} L_{\odot}$.

The shape of both spectra is similar to that of the massive protostar
candidate IRAS 23385+6053 (Fontani et al. \cite{fonta2}). In addition,
as pointed out in Sect.~\ref{cont}, the source structures are also similar.
This may suggest that the evolutionary stage of the sources is the
same.
We will come back to this point in Sect.~\ref{nature}.

\subsection{Mass, density, and temperature estimates}
\label{mass}
Core masses have been derived in two different ways: from the continuum
emission and from the lines. In the first case, assuming dust emission
to be optically thin, we can compute the core mass $M$ directly from the
observed flux density:
\begin{equation}
M=\frac{F_{\nu}d^{2}}{k_{\nu}B(\nu,T)} \label{emcont}
\end{equation}
where $F_{\nu}$ is the flux density integrated over the area described by
the $3 \sigma$ level of the SCUBA map, $d$ is the distance, 
$k_{\nu}$ is the dust opacity parametrised as
$k_{\nu}=k_{\nu_{0}}(\nu/230\;{\rm GHz})^{\beta}$ (in units of cm$^{2}$ g$^{-1}$),
$B(\nu,T)$ is the Planck function and $T$ is the dust temperature. We
assume $k_{\nu}=0.005$ cm$^{2}$ g$^{-1}$ (Kramer et al.~\cite{kramer}), that
implies a standard gas-to-dust ratio of 100. We have also deduced the core mass 
by means of the total column density derived from some of the lines observed,
according to the expression:
\begin{equation}
M_{\rm line}=\frac{\pi}{4} \frac{D^{2} N_{\rm tot}m_{\rm H_{2}}}{X}     
\label{emcd}
\end{equation}
where $D$ is the source diameter, $m_{\rm H_2}$ the mass of the H$_2$
molecule, $X$ the abundance of the molecule relative to H$_2$,
and $N_{\rm tot}$ is the source averaged column density.
For the \CII\ molecule, following Wilson \& Rood 
(\cite{wilson}) we assumed that the oxygen isotope ratio $^{16}$O/$^{18}$O
depends on the galactocentric distance $R_{\rm GC}$, according to the 
relationship
${\rm ^{16}O/^{18}O}=58.8 R_{\rm GC}({\rm kpc})+ 37.1$; using the 
galactocentric distances given in
Table~\ref{tsources}, and assuming a 
mean abundance of $10^{-4}$ for the \CO\ molecule, we infer a \CII\ mean
abundance of $\sim1.9\times 10^{-7}$.
For the \HCOpI\ molecule, we applied a similar argument: the carbon 
isotope ratio, $^{13}$C/$^{12}$C, 
depends on the galactocentric distance $R_{\rm GC}$, according to the 
relationship
${\rm ^{12}C/^{13}C}=7.5 R_{\rm GC}({\rm kpc})+ 7.6$ ; thus, assuming a 
mean abundance 
of $X=10^{-9}$ for the \HCOp\ molecule (Irvine et al. \cite{irvine}), we 
derive a \HCOpI\ mean abundance of $\sim 1.5\times 10^{-11}$. 

Furthermore, from the linewidth and the core linear diameter of each 
molecular line, we can estimate the virial mass: 
assuming the source to be spherical and homogeneous, neglecting contributions
from magnetic field and surface pressure, the virial mass is given by:
\begin{equation}
M_{\rm VIR}(M_\odot)=0.509\,d({\rm kpc})\,\Theta_{\rm s}({\rm arcsec})\,\Delta v_{1/2}^{2}({\rm km/s})   \label{emvir}
\end{equation}

The mass estimates deduced from Eqs.~(\ref{emcont}) and (\ref{emcd}) are
listed in Col. 5 of Tables~\ref{tm136_core} and \ref{tm143_core},
while the virial masses are shown in Col. 4. 
In Col. 6 we also give the mean $\rm H_{2}$ volume density, $n_{H_{2}}$, 
deduced from the corresponding mass and diameter. 
The values found clearly show that both cores are massive, and
the masses needed for virial equilibrium are within a 
factor of $\sim2$ from those computed from
the continuum emission or from the lines. Hence the cores are likely 
virialised. Furthermore, $n_{\rm H_{2}}$ increases towards the core center in
both sources; the increment is roughly consistent with what is expected for a 
density profile of the type $n_{\rm H_{2}}\propto R^{-2}$, which is the typical
profile found in molecular clumps that surround high-mass YSOs (see e.g.
Fontani et al. 2002, Beuther et al. \cite{beuther2}). As usual, the 
uncertainties on the mass estimates are mainly due to the molecular abundances,
the assumed dust opacity and the gas-to-dust ratio, which are known to be
affected by large errors and difficult to quantify. 

One notices that the masses estimated from the 
flux densities at 850 $\mu$m are larger than those derived from the
\CII\ (2--1) line, although the source diameters are bigger 
in the \CII\ (2--1) line. This may be due to the 
fact that all diameters have been estimated from the FWHM of the
emission, and in the 850 $\mu$m maps a significant fraction of the total
integrated emission
arises from a halo of $\sim$40\asec\ in diameter, which is 
not detected in the line maps.

Finally, from the \ace\ spectra observed towards \MII\ (Fig.~\ref{spec_ace}), 
we have derived the kinetic temperature and the total column density of this 
source by means of the rotational diagram method (see e.g. Fontani et al
\cite{fonta}). The fundamental assumption of this method is that the gas is
in local thermodynamic equilibrium (LTE) conditions, which we believe to
be valid for \ace\ because of its low dipole moment. If we further
assume the lines to be optically thin, one can derive the column density
$N_{\rm i}$ of the upper level of each transition from the integrated 
intensity of the line. We obtain the source kinetic temperature 
and column density listed in Cols. 1 and 2 of Table~\ref{tm143_temp}, 
respectively.
We also list the virial mass (Col. 3), the gas mass (Col. 4)
and the $\rm H_{2}$ volume density (Col. 5), computed as explained before for 
the other
molecular tracers. The diameter used to compute $n_{\rm H_{2}}$ is that
derived from the SCUBA map. We assumed an abundance of \ace\
relative to H$_{2}$ of $2\times10^{-9}$. This is an average value taken
from Fontani et al. (\cite{fonta}) and Wang et al. (\cite{wang}), who
studied this molecule in massive star forming regions.

\begin{table}
\begin{center}
\caption[] {Temperature, mass and ${\rm H_2}$ volume density estimates obtained
from CH$_{3}$CCH for \MII . A \ace\ abundance of $2\times10^{-9}$ has been
assumed. As source diameter, we have used that of the SCUBA map.}
\label{tm143_temp}
\begin{tabular}{ccccc}
\hline
 $T_{\rm k}$ & $N_{\rm tot}$ & $M_{\rm vir}$ & $M_{\rm H_2}$ & $n_{\rm H_{2}}$ \\
 (K) & (cm$^{-2}$) & ($M_\odot$) & ($M_\odot$) & (cm$^{-3}$) \\
\hline
 27  &  4 10$^{13}$ & 73  & 92 & 1.0 10$^{5}$ \\
\hline
\end{tabular}
\end{center}
\end{table}

\subsection{The outflows}
\label{outflow}

\subsubsection{Integrated maps and morphology}
\label{morf}

\begin{figure}
\centerline{\includegraphics[angle=0,width=7.5cm]{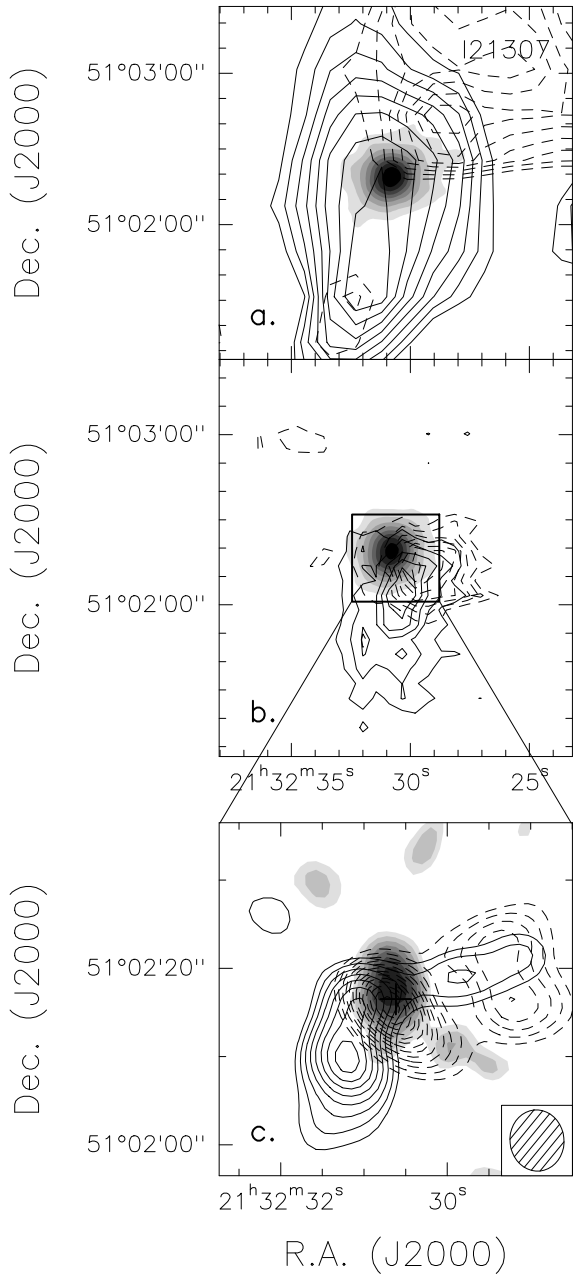}}
\caption{Molecular outflow observed towards \MI .
{\bf a}: \CO\ (2--1) line observed with the
NRAO-12m telescope, integrated under the blue-shifted (solid line)
and the red-shifted (dashed line) wings. The velocity ranges are $(-55, -50)$
\kms\ (blue) and $(-44,-38)$ \kms\ (red). Contour levels begin
from 6 ($\sim$4$\sigma$) K \kms , with a spacing of 2 K \kms . 
The overlaid grey-scale represents
the SCUBA 850 $\mu$m emission map: the same contour levels as in 
Fig.~\ref{m136_map_tot_isocam} are shown.
{\bf b}: Same as {\bf a} for the \CO\ (2--1) line observed with
the IRAM-30m telescope. Contour levels begin
from 2.5 K \kms\ ($\sim 3\sigma$ level in the maps) and have a spacing of 2 
K \kms .
{\bf c}: Same as {\bf a} and for the \CO\ (1--0) line observed with OVRO. 
Contour values begin from 0.9 K \kms\
($\sim3\sigma$) and have a spacing of 0.3 K \kms . The 
grey-scale represents the core I21307-C observed with OVRO at 3~mm, with the 
same contours as in Fig.~\ref{m136_map_tot_isocam}.}
\label{scuba_nrao_m136}
\end{figure}

\begin{figure}
\centerline{\includegraphics[angle=0,width=7.5cm]{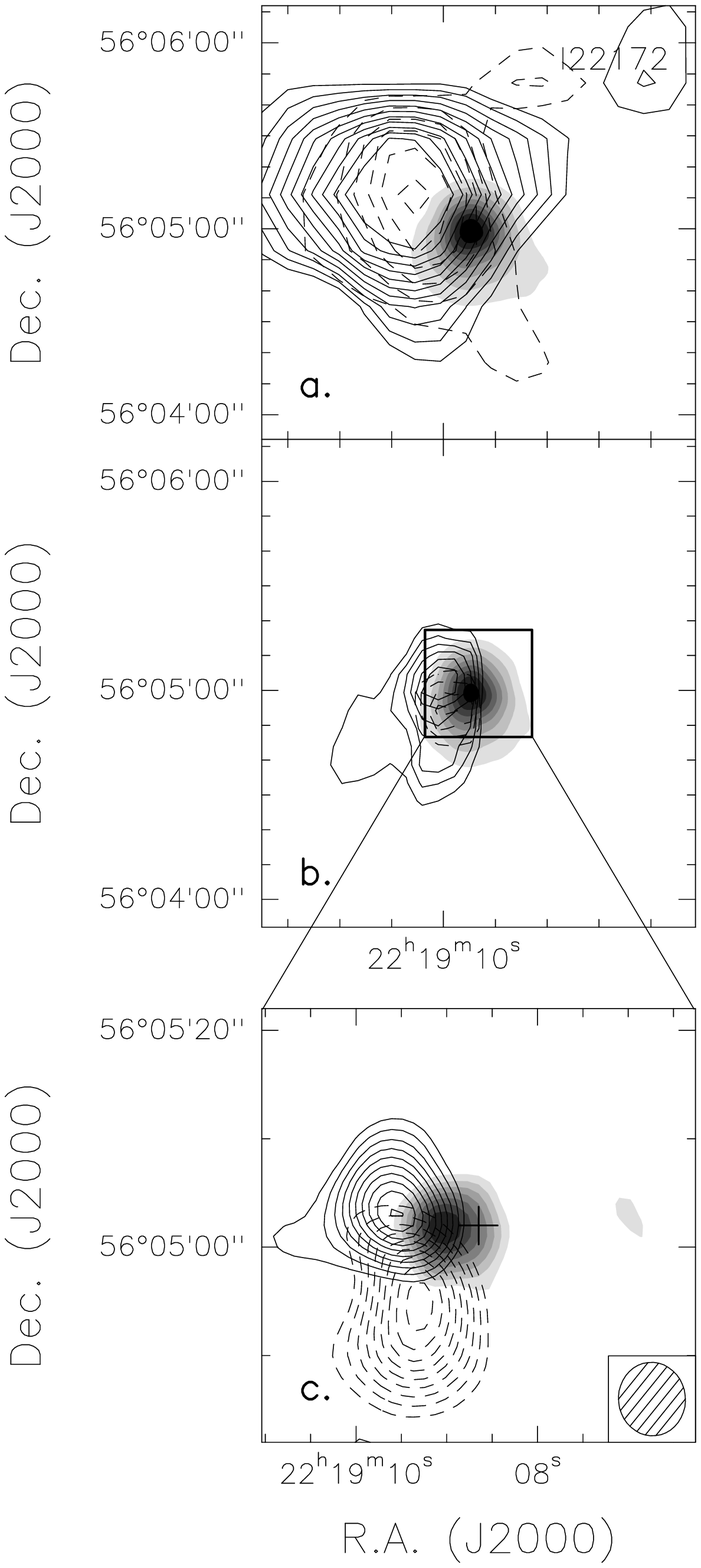}}
\caption{Same as Fig.~\ref{scuba_nrao_m136} for \MII . 
{\bf a}: contour levels of the blue- and red- wings begin from 8 
($\sim$4$\sigma$) K \kms , with a 
spacing of 4 K \kms. The velocity ranges are $(-51,-45)$
\kms\ (blue) and $(-38,-30)$ \kms\ (red). The grey-scale represents
the SCUBA 850 $\mu$m emission map: the same contour levels as in 
Fig.~\ref{m143_map_tot} are shown.
{\bf b}: contour levels begin from 5 K \kms\ ($\sim 3\sigma$ level 
in the maps) and have a spacing of 5 K \kms .
{\bf c}: same as {\bf c.} of Fig.~\ref{scuba_nrao_m136}. Contour levels 
of the blue and red wings begin from 1.2 K \kms , and have a spacing of 
0.4 K \kms . The grey-scale represents the OVRO 3~mm map, identifying the 
core I22172-C, with the same contours as in Fig.~\ref{m143_map_tot}.}
\label{scuba_nrao_m143}
\end{figure}

Figures~\ref{scuba_nrao_m136} and \ref{scuba_nrao_m143} show the integrated
intensity maps of the red and the blue wings of the \CO\ (2--1) line. Panels 
{\bf a} and {\bf b} show the observations performed  
with the NRAO-12m telescope and the IRAM-30m telescope, respectively. 
Solid 
and dashed lines correspond to the integrated emission under the blue and 
the red wings, respectively. The velocity intervals are given in 
Table~\ref{tline}. These have been
superimposed on the SCUBA map at 850 $\mu$m, which indicates the position 
of the cores. 

The outflow associated with \MI\ is oriented
approximately NW-SE, with the blue-shifted gas towards the SE, and the lobes
appear to be well separated. In contrast, towards 
\MII\ the lobes overlap completely. However, in
both cases, the angular resolution is too poor to resolve the outflow 
structure. The interferometric observations obtained with OVRO 
in the \CO\ (1--0) line give us better understanding of the structure of the 
flows: in panels {\bf c} of Figs.~\ref{scuba_nrao_m136} and 
\ref{scuba_nrao_m143}, the outflow maps are shown. 
These have been obtained integrating the line emission over the same velocity
intervals as in panels {\bf a} and {\bf b}.
 
Let us consider the two outflows separately.
\MI\ (Fig.~\ref{scuba_nrao_m136}{\bf c}) presents an outflow oriented NW-SE, 
with 
the blue-shifted gas located to the SE. The lobe peaks are centered near the
core position. The morphology is quite chaotic
and the outflow seems to be made at least of two components: a stronger one 
centered on the core I21307-C, and a weaker one, offset by $\sim15$\asec\ 
to the west. 
Various authors (e.g. Beuther et al. \cite{beuther}) 
have demonstrated the presence of multiple outflows in massive star 
forming regions. Hence, such a complex morphology may be due to the 
fact that we are seeing the superimposition of several flows.

In \MII\ (Fig.~\ref{scuba_nrao_m143}{\bf c}), the outflow
axis is clearly oriented N-S, with red-shifted gas to the south and 
blue-shifted gas to the north.
The whole outflow has an extension of $\sim30$\asec ,
corresponding to $\sim0.35$ pc at 2.4 kpc.
In this case the lobes have similar sizes and there is little overlap 
between them. 
It is striking that the center of the lobes is offset by 
$\sim 5$\asec\ from the dust emission peak. This suggests that the 
source driving the outflow is not located at the center
of the dust emission. The outflow might originate
from a source belonging to the more evolved cluster, and not from the 
massive object embedded inside I22172-C; in fact, comparing
the image at 1.7$\mu$m (bottom panel of Fig.~\ref{m143_nir}) with the  
panel {\bf c} of Fig.~\ref{scuba_nrao_m143}, one notices that the center 
of the lobes roughly corresponds to one of the near-infrared sources.
Thus, the molecular core might  
host a source in a very early evolutionary stage prior to the formation 
of an outflow. However, as we will briefly discuss in 
Sect.~\ref{phipar}, the flow
parameters indicate that the source driving the outflow must be massive.
Therefore, wherever the source driving the outflow is located, it must
be a massive young object. Since the stars of the cluster are probably
low- to intermediate-mass stars, it is unlikely that the flow originates in
one of the cluster members.
 
We can compare our OVRO maps with those presented in Molinari et al. 
(\cite{mol02}), who used the \HCOp\ (1--0) line to observe the
outflows in both sources. They 
adopted velocity intervals slightly different from ours. After
integrating the \CO\ (1--0) lines in
the same intervals, one sees that our maps are consistent 
with those of Molinari et al. (\cite{mol02}) for \MI , but not for \MII . 
This might be explained by the fact that 
Molinari et al. (\cite{mol02}) observed only the inner part of the
region observed by us using a more compact configuration of 
the interferometer, sensitive to more extended structures.

\subsubsection{Physical parameters}
\label{phipar}

The characteristics of the flows are listed in Table~\ref{tm136_out},
in which the source and the tracer used are given in Col. 1 and 3, 
respectively. The derived physical properties have been computed 
using the same relations given in Beuther et al.\cite{beuther}: the
mean column densities of the blue- and the red- lobes, $N_{\rm b}$ and 
$N_{\rm r}$, are given in Col.~4 and 5, the mass of the blue- and the 
red- lobes $M_{\rm b}$ and
$M_{\rm r}$, and the total mass, $M_{\rm out}$, in Col. 6, 7 and 8, 
respectively, the momentum $P_{\rm out}$ in Col.~9, 
the energy $E$ in Col.~10, the size $r$ in Col.~11, the dynamical timescale
$t_{\rm dyn}$ in Col.~12, the 
mass entrainment rate $\dot{M}_{\rm out}$ in Col.~13, the mechanical force 
$\dot{P}_{\rm out}$ in Col.~14 and the mechanical luminosity $L_{\rm m}$ 
in Col.~15. 
All values have been derived assuming the mean inclination angle of
$57^{\circ}$ (Cabrit \& Bertout~\cite{cabrit2}) with respect to the line 
of sight.

To determine the size
$r$, we have adopted the maximum extension from the center of the lobes 
(Cabrit \& Bertout \cite{cabrit2}). Since in \MI\ the lobes have 
different extensions, we have adopted their arithmetic mean. 

\begin{table*}
\begin{center}
\caption[] {Physical parameters of the outflows.
Column densities of the blue- and the red- wing ($N_{\rm b}$ and $N_{\rm r}$)
are expressed in units of $10^{21}$cm$^{-2}$, masses ($M_{\rm b}$, $M_{\rm r}$
and $M_{\rm out}$) in $M_{\odot}$, momentum ($P_{\rm out}$) in 
M$_{\odot}$ km s$^{-1}$, 
energy ($E$) in $10^{45}$ ergs, size ($r$) in pc, dynamical timescale ($t_{\rm dyn}$) in $10^{4}$ yr, mass entrainment rate ($\dot{M}_{\rm out}$) in $10^{-4}M_{\odot}$ yr$^{-1}$, mechanical force ($\dot{P}_{\rm out}$) in 
$10^{-3}M_{\odot}$\kms\ yr$^{-1}$
and mechanical luminosity ($L_{\rm m}$) in $L_{\odot}$. All values
have been calculated assuming the mean inclination angle of 
$57^{\circ}$.}
\label{tm136_out}
\begin{tabular}{cccccccccccccc}
\hline
   Source & tracer & $N_{\rm b}$ & $N_{\rm r}$ & $M_{\rm b}$ & $M_{\rm r}$ &
$M_{\rm out}$ & $P_{\rm out}$ & $E$ & $r$ &$t_{\rm dyn}$ & $\dot{M}_{\rm out}$& $\dot{P}_{\rm out}$ & $L_{\rm m}$  \\
\hline
\MI\ & \CO\ (1--0) (OVRO) & 33.8 & 17.9 & 2.3 & 0.8 & 3.2 & 48.3 &  3.7 &  0.1 &  1.3 &
 4.5 &  2.9 &  10.5  \\
     & \CO\ (2--1) (30m) & 0.8 & 0.7 & 5.2 & 3.0 & 8.2 &  127.8 &  20.0  &  0.7 &  4.3 &  1.9 &  2.9 &  6.8 \\
     & \CO\ (2--1)(12m) & 0.5 & 0.4 & 15.8 & 4.2 & 20.0 &  311.0 &  48.3 & 1.0 &  5.7 &  3.4  &  5.5  &  13.1 \\
\MII\ & \CO\ (1--0) (OVRO) & 52.4 & 39.1 & 1.1 & 1.6 & 2.7 &  55.6 &  12.4 &  0.1 &  0.7 &  7.6 &  12.2 &  14.2 \\
    & \CO\ (2--1) (30m) & 1.2 & 1.6 & 5.0 & 3.4 & 8.4 &  153.5 &  31.0 &  0.4 &  1.7 &  4.7 &  9.0 &  27.3 \\
    & \CO\ (2--1)(12m) &  1.0 & 1.1 & 12.8 & 10.1 & 22.9 &  428.7 &  89.3 &  0.6 &  2.9 &  7.8 &  14.8 &  79.3 \\ 
\hline
\end{tabular}
\end{center} 
\end{table*}

The velocity intervals chosen for the line wings are the same used for the
OVRO \CO\ (1--0) spectra (see Table~\ref{tline}). 

The \CO\ (2--1) and (1--0) lines may have large 
opacities also in the wings. Hence, we should take into account 
this effect in computing the column density. For the \CO\ (2--1) line we 
have followed the method outlined in Beuther 
et al. (\cite{beuther}): one can assume a mean value for the 
$^{13}$CO/$^{12}$CO (2--1) 
line wing ratio of $\sim 0.1$ (Choi et al. \cite{choi}), and that this 
ratio is constant throughout the outflow (Cabrit \& Bertout \cite{cabrit}). 
Even though we have not
made maps of the $^{13}$CO (2--1) line, we can verify if this 
assumption holds for
our sources using the data of 
Brand et al. (\cite{brand}): they
observed the $^{13}$CO (2--1) line towards \MI , and by comparing their spectra
with ours in the \CO\ (2--1) line, we conclude that a mean ratio of 0.1 is 
plausible for the wings.
Under this assumption, the ${\rm H_{2}}$ column densities of the wings, 
$N_{\rm b}$ and $N_{\rm r}$, have 
been computed using the relation (see Beuther et al. \cite{beuther}):
\begin{eqnarray}
N&=&\left(\frac{\rm H_{2}}{\rm ^{13}CO}\right)\frac{3k^{2}T_{\rm ex}}{4\pi^{3}h\nu^{2}\mu^{2}}\exp(-16.6/T_{\rm ex}) \nonumber \\  
& &0.1\int_{wing} T_{\rm MB}\;^{12}{\rm CO}(2-1){\rm d}v\;, 
\label{col}
\end{eqnarray}
We have assumed $\frac{\rm H_{2}}{\rm ^{13}CO}=89\times 10^{4}$ and 
$T_{\rm ex}=30$ K as in Beuther et al. (\cite{beuther}), to make a consistent 
comparison. For this reason the correction for the galactocentric distance 
has not been applied. However, such a correction would not affect any of the
conclusions drawn by us and Beuther et al..

For the \CO\ (1--0) line, we have estimated the optical depth in the wings
as follows: from the mean line wing ratio $^{13}$CO/$^{12}$CO (2--1)$=0.1$,
and a relative mean abundance
$\frac{\rm H_{2}}{\rm ^{13}CO}=89\times 10^{4}$,
we derive an optical depth of the \CO\ (2--1) line in the wings of  
$\tau\sim5$. One can demonstrate that, in LTE
conditions, for a kinetic temperature of 30 K, the optical depth of the 
(2--1) line is $\sim$3 times that of the (1--0) line. Hence, we adopted
a mean value $\tau=1.7$ for the \CO\ (1--0) line to compute the column
density in the wings, and corrected the column density estimates accordingly. 

The parameters deduced from the \CO\ (1--0) line are in good agreement with
those previously found by Molinari et al. (\cite{mol02}), who observed 
the \HCOp\ (1--0) line towards both sources with similar angular resolution,
as mentioned in Sect.~\ref{morf}. The outflow masses, the mechanical forces
and the mechanical luminosities are at least an order 
of magnitude greater than those typically associated with low-mass YSOs 
(see e.g. Bontemps et al.~\cite{bontemps}), and 
in good agreement with those found by Beuther et al. (\cite{beuther}) in
a sample of massive YSOs, believed to be precursors
of UC \HII\ regions. A detailed comparison between our results and 
those found by other authors (both in low-mass and high-mass star forming 
regions) will be done in Sect.~\ref{comparison}.

\subsubsection{Near-Infrared H$_{2}$ and [FeII] images.}

\begin{figure}
\centerline{\includegraphics[angle=0,width=8.cm]{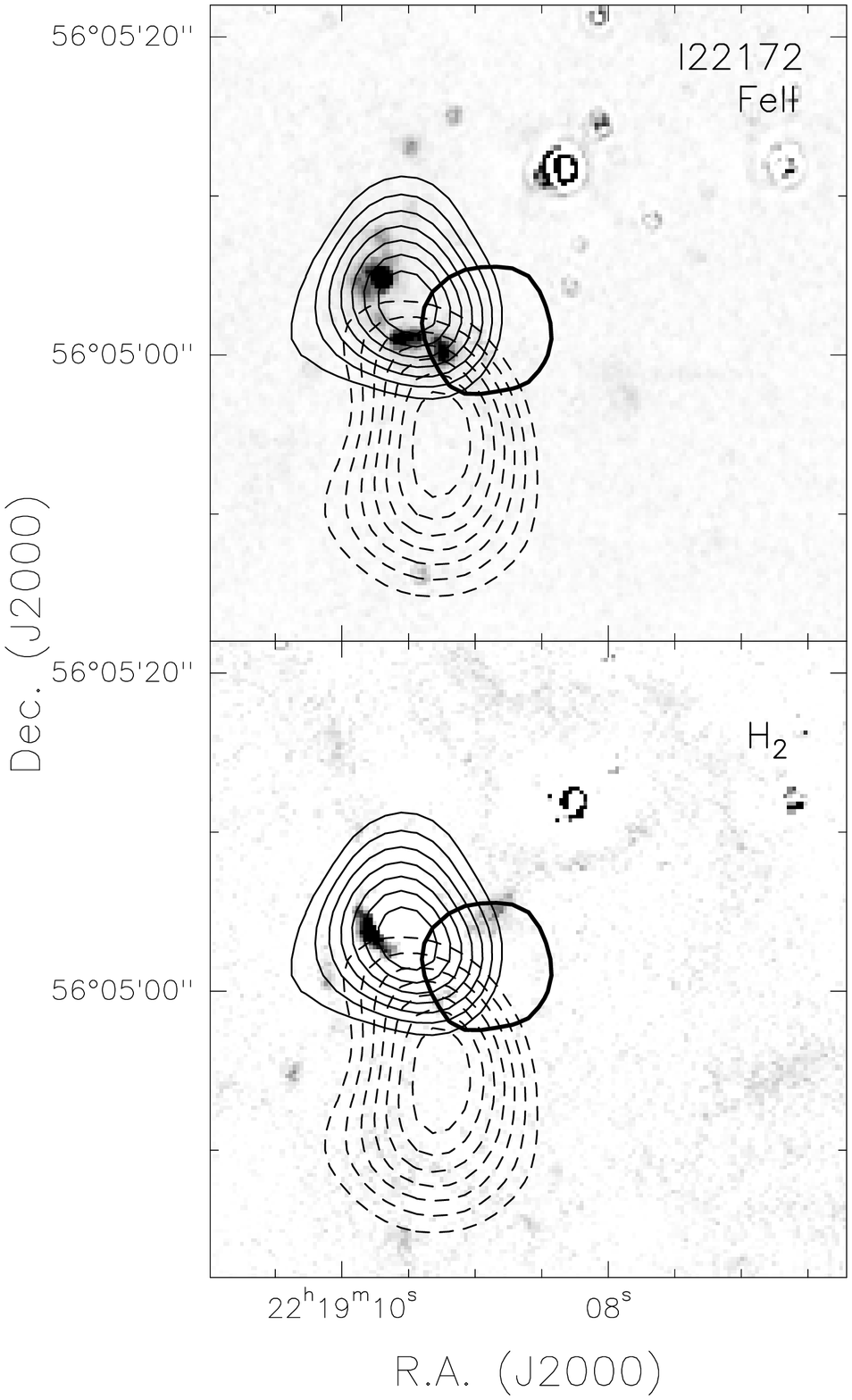}}
\caption{[FeII] and H$_{2}$ line images of \MII\ after continuum subtraction. 
The solid
and dashed lines indicated the blue and red lobes of the outflow,
respectively, traced by the \CO\ (1--0) line. The same contour levels as in
Fig.~\ref{scuba_nrao_m143} are shown. The position of the core I22172-C is
indicated by the thick contour, which represents the half maximum power 
level of the OVRO map.}
\label{jet}
\end{figure}

Fig.~\ref{jet} presents the image of \MII\ in the H$_{2}$ 
 and [FeII] lines. They have been superimposed on the outflow image 
derived from the \CO\ (1--0) line. Also indicated in Fig.~\ref{jet} is
the position of the core I22172-C.
The [FeII] emission is in the center
of the blue lobe of the outflow, and it is oriented approximately NE-SW. 
The H$_{2}$ emission has roughly the same inclination, and
it seems to follow the boundary of the [FeII] emission. 
Hence, we conclude that both these tracers are associated with
the \CO\ flow seen with OVRO.

It is well known
that NIR lines of H$_{2}$ and [FeII] are tracers of the hotter and faster
component of a flow in low-mass YSOs (e.g. Nisini et al. \cite{nisini}).
In particular, 
[FeII] is thought to be a tracer of the inner region of the jet, while the 
H$_{2}$ emission is
likely due to the shocked gas on the surface between the jet and the 
ambient material. From Fig.~\ref{jet} one can see that our observations are 
consistent with this picture: the H$_{2}$ and [FeII] emission both arises 
from the inner parts of the outflow lobes.

No H$_2$ or [FeII] line emission was detected in \MI .

\section{Discussion}
\label{discu}

In order to understand the nature of IRAS 21307+5049 and 
IRAS 22172+5549, in the following we will discuss the results obtained in 
Sect.~\ref{res}.

\subsection{Comparison with outflows associated with other YSOs}
\label{comparison}

\begin{figure}
\centerline{\includegraphics[angle=-90,width=8.cm]{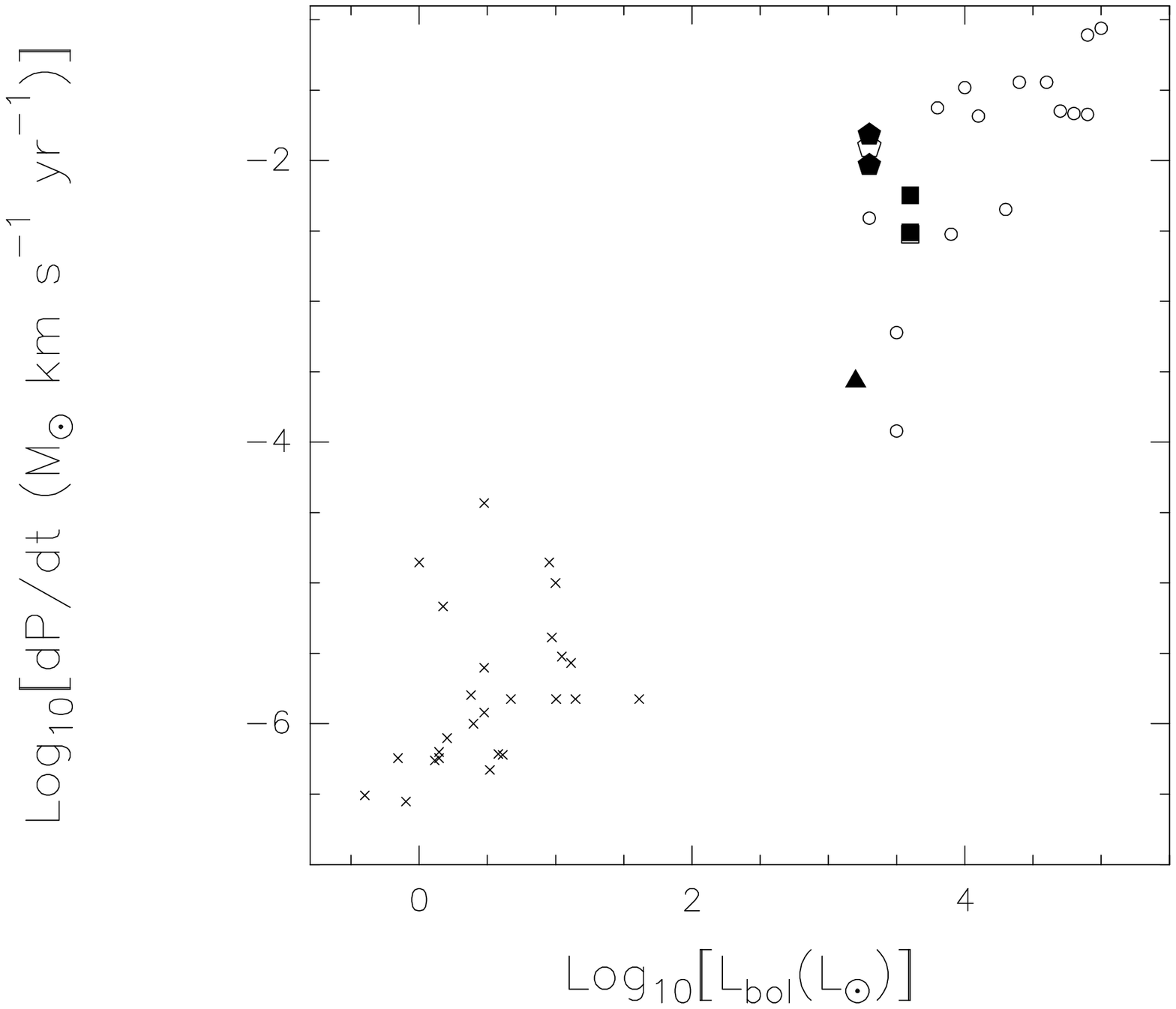}}
\caption{Mechanical force of the outflows, $\dot{P}_{\rm out}$, versus 
bolometric luminosity $L_{\rm bol}$ of the source. \MI\ and \MII\ data are 
represented by squares and
pentagons, respectively (open symbol = OVRO data, filled symbols = 
single-dish data); crosses indicates data from Bontemps et al. 
(\cite{bontemps}) (low-mass YSOs) and open circles correpond to those of
Beuther et al. (\cite{beuther}) (high-mass YSOs). The filled triangle indicates
IRAS 23385+6053 (Fontani et al. \cite{fonta2}). }
\label{p_l}
\end{figure}

In Fig.~\ref{p_l}, we plot the mechanical force, $\dot{P}_{\rm out}$, of the 
outflows against the
bolometric luminosities $L_{\rm bol}$ of 
the sources. Our data are compared to those derived by Bontemps et al. 
(\cite{bontemps}), who studied outflows associated with low-mass objects,
and those of Beuther et al. (\cite{beuther}), who analysed massive
molecular outflows. 
We also compare our results to those obtained by 
Molinari et al. (\cite{mol98b}) and Fontani et al. (\cite{fonta2}) for
IRAS 23385+6053, a candidate massive protostar.
Figure~\ref{p_l} is similar to Fig.~4b of Beuther et al. (\cite{beuther}); 
however, unlike these
authors, we consider only sources without 
ambiguity between the near and the far kinematical distance.
Also the values computed by Beuther et al.~(\cite{beuther})
have been corrected for the mean inclination angle of $57^{\circ}$:
this allow us to make a consistent comparison.

Beuther et al. (\cite{beuther}) have shown that there is a correlation between
the mechanical force $\dot{P}_{\rm out}$ and the luminosity of the
core driving the outflow. They suggest that a similar relation,
established for low-mass objects 
(Bontemps et al. \cite{bontemps}), holds up to the high-mass regime. Our 
results and those obtained by Fontani et al.~(\cite{fonta2}) for 
IRAS 23385+6053 appear to confirm this statement.
In particular, \MI\ and \MII\ have bolometric luminosities close to 
those of early B stars.

Finally, we have computed the dynamical timescale of the outflows.
These values represent a lower limit to the outflow age
(see e.g. Parker~\cite{parker}). For \MI\ and \MII , we have found 
(see Table~\ref{tm136_out}) that 
$t_{\rm dyn}$ is of the order of $10^{4}-10^{5}$ yr.
Beuther et al. (\cite{beuther}) found 
values ranging from $2\times10^{4}$ to $2\times10^{5}$ yr.
These values are again in good agreement with our estimates.

\subsubsection{Mass accretion rates and accretion efficiencies}
\label{macc}

From the mass entrainment rate, we can derive the mass accretion rate,
$\dot{M}_{\rm acc}$, and the accretion efficiency, $f_{\rm acc}$.
For both, we will adopt the same argument outlined by Beuther et 
al. (\cite{beuther}).

Based on the assumptions that the source driving the outflow is a single star,
the momentum of the outflow is equal to that of the internal jet and
that the ratio between mass loss rate of the jet and mass accretion rate
is $\dot{M}_{\rm jet}\simeq 0.3\dot{M}_{\rm acc}$ (Tomisaka~\cite{tomi};
Shu et al.~\cite{shu99}), we obtain accretion rates
of order $10^{-4}M_{\odot}{\rm yr^{-1}}$ (see Table~\ref{tacc}). 
Such values are slightly higher than those derived by Beuther 
et al. (\cite{beuther}) in sources with comparable luminosities 
(from $\sim10^{3}$ to $\sim10^{4}\;L_{\odot}$), but
still consistent within the uncertainties.

From the accretion rate we estimate the accretion efficiency, 
$f_{\rm acc}$:
\begin{equation}
f_{\rm acc}=\frac{\dot{M}_{\rm acc}}{M_{\rm core}/t_{\rm ff}}\;.
\label{efacc}
\end{equation}
Using $\dot{M}_{\rm acc}$, $t_{\rm ff}$, and the values of
$M_{\rm core}$ computed from the 850 $\mu$m map, we derive
$f_{\rm acc}\sim0.03$ and $\sim0.08$ for \MI\ and \MII , respectively  
(see Table~\ref{tacc}). 

$f_{\rm acc}$ can also be determined following the 
argument outlined in Richer et al. (\cite{richer}). They consider two 
approaches:
\begin{itemize}
\item [(a)]the luminosity is accretion dominated:
\begin{equation}
f_{\rm acc}=2\times10^{-3}\frac{L}{\dot{P}_{\rm out}v_{\rm kep}}
\frac{v_{\rm jet}}{v_{\rm kep}}
\label{fr_acc}
\end{equation}
\item [(b)]the luminosity is mainly due to ZAMS stars: 
\cite{beuther}):
\begin{equation}
f_{\rm acc}=2\times10^{-3}\frac{M_{*}v_{\rm kep}}{P_{\rm out}}
\label{fr_zams}
\end{equation}
\end{itemize}
where $v_{\rm kep}=\sqrt{GM_{*}/R_{*}}$ is the stellar escape velocity,
assumed equal to the jet velocity, and
$M_{*}$ and $R_{*}$ are the stellar mass and radius for a ZAMS star with 
bolometric luminosity $L$. The luminosities used are those derived in
Sect.~\ref{lum}. $P_{\rm out}$ and $\dot{P}_{\rm out}$ have been 
calculated as explained in Sect.~\ref{phipar}. 
\begin{table*}
\begin{center}
\caption[] {Accretion rates and accretion efficiencies}
\label{tacc}
\begin{tabular}{cccccc}
\hline \hline
source & tracer & ${\dot M_{\rm acc}}$ &  $f_{\rm acc}$ (*) & 
$f_{\rm acc}$ (a) &  $f_{\rm acc}$ (b) \\
       &        & ($10^{-4}\;M_{\odot}{\rm yr^{-1}}$)    &               &         &       \\
\hline
\MI\ & \CO\ (1--0) &         0.8 &  0.04  & 0.01 & 0.4 \\
     & \CO\ (2--1) (30-m) &  0.6 &  0.03 & 0.01 & 0.2 \\
     &  \CO\ (2--1) (12-m) & 0.4 & 0.02 & 0.02 & 0.2 \\
\MII\ & \CO\ (1--0) &        1.3 & 0.07 & 0.003 & 0.3 \\
      & \CO\ (2--1) (30-m) & 1.4 & 0.08 & 0.003 &  0.1 \\
      & \CO\ (2--1) (12-m) & 0.8 & 0.05 & 0.006 &  0.09 \\
IRAS 23385+6053 ($^{\clubsuit}$)& \HCOp\ (1--0) & 1.5 & 0.006 & 0.01 & 0.1 \\
Beuther et al. ($^{\spadesuit}$) &\CO\ (2--1) (30-m) & 0.3 
 & 0.02 & 0.08 & 0.2 \\
\hline
\end{tabular}
\end{center}
($^{\clubsuit}$) Molinari et al. (\cite{mol98b}) \\
($^{\spadesuit}$) average values for the sources of the Beuther et al. 
(\cite{beuther}) sample with luminosity from $10^{3}$ to $10^{4}\;L_{\odot}$\\
(*) computed from Eq.~(\ref{efacc}) \\
(a) computed from Eq.~(\ref{fr_acc})  \\
(b) computed from Eq.~(\ref{fr_zams}) \\
\end{table*}
 
The values of $f_{\rm acc}$ computed from Eqs.~(\ref{fr_acc}) and 
(\ref{fr_zams}) are listed 
in Table~\ref{tacc}, together with the result for
IRAS 23385+6053 and the average values
deduced for the sources of the Beuther et al. sample with comparable
luminosity.  

\subsection{Nature of the sources}
\label{nature}

\begin{figure}
\centerline{\includegraphics[angle=-90,width=8.cm]{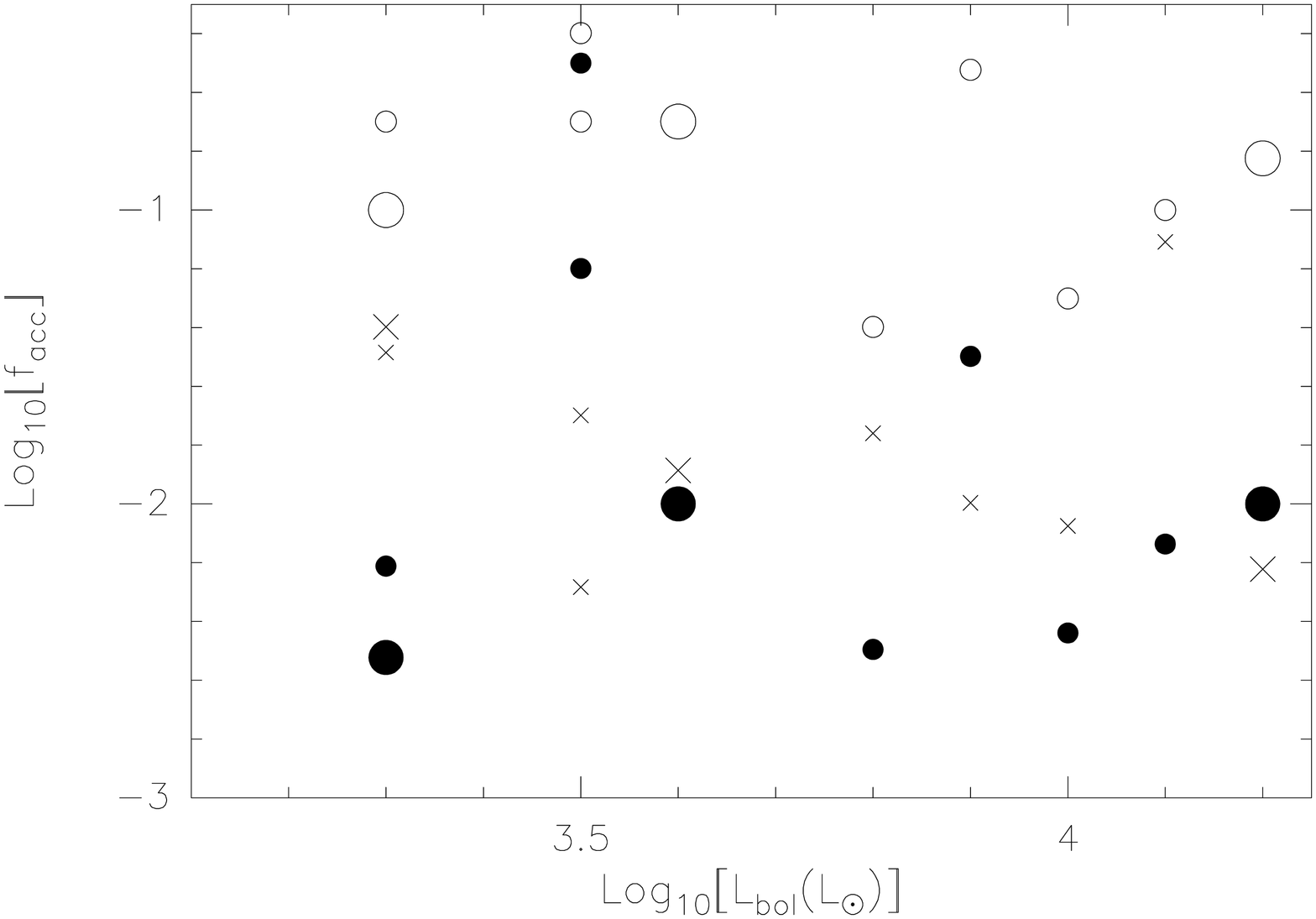}}
\caption{Accretion efficiencies estimated from Eq.~(\ref{efacc}) (crosses),
from Eq.~(\ref{fr_acc}) (filled circles) and from Eq.~(\ref{fr_zams}) (open
circles), against the bolometric luminosity. Big symbols correspond to
\MI , \MII\ and IRAS 23385+6053; small symbols indicate the sources of the
Beuther et al. (\cite{beuther}) sample.}
\label{ffacc}
\end{figure}
We stress that the quantities previously obtained have been derived 
under several hypotheses, the most important being the assumed jet velocity 
and the efficiency of the ejection mechanism (i.e. the ratio between mass
loss rate and accretion rate, assumed to be 0.3), which are both very 
uncertain. 
Hence, the values listed in Table~\ref{tacc} have to be taken with
caution. Neverthless,
the results presented in Sects.~\ref{comparison}
and \ref{macc} indicate that
\MI\ and \MII\ have physical parameters comparable to those obtained by 
Beuther et al. (\cite{beuther}) in sources with similar luminosities, 
and hence likely in similar evolutionary stages.

In Fig.~\ref{ffacc} we plot $f_{\rm acc}$ estimated
using Eq.~(\ref{efacc}), (\ref{fr_acc}) and (\ref{fr_zams}) for \MI ,
\MII\ and IRAS 23385+6053 (big symbols), and for the 
sources of the Beuther et al. sample with comparable luminosity (small
symbols). Although the spread of the values
is about an order of magnitude for the three estimates of $f_{\rm acc}$, one 
can notice that the values
derived from Eq.~(\ref{efacc}) are more consistent with those computed
from Eq.~(\ref{fr_acc}), i.e. under the assumption of accretion dominated 
luminosity, than with
those estimated in the case of luminosity dominated by a ZAMS star.
This result favours the conclusion that our sources, as well as those of
Beuther et al., are likely massive protostars still in the accretion phase. 

\subsubsection{Comparison with the massive protostar candidate IRAS 23385+6053}
\label{m160}

In Sects.~\ref{cont} and~\ref{lum} we have shown that the continuum maps
indicate the presence of two ``sub-regions'' in
both \MI\ and \MII : a massive compact core (called I21307-C and
I22172-C)
and a cluster of stars which surround the core. From the maps and the SEDs, 
we concluded that the core is responsible
for the continuum emission at wavelengths longer than $\sim 25\;\mu$m,
while the emission at shorter wavelengths is mainly due to the  
cluster. Grey-body fits to the SEDs indicate that the cores
have luminosities of $\sim10^{3}\;L_{\odot}$, an order of magnitude 
greater than those of the clusters.

The structure of both objects is similar to that of 
IRAS 23385+6053. This source, which has a bolometric luminosity of 
$\sim 10^{4}\;L_{\odot}$, has been proposed by Molinari et al. 
(\cite{mol98b}) and Fontani et 
al. (\cite{fonta2}) as the best example of a candidate massive protostar. 
As in \MI\ and \MII , we can distinguish two components: a compact 
core detected at millimeter and sub-millimeter wavelengths and a cluster of 
stars which lie around the core, detected at near- and mid-infrared
wavelengths. Other important similarities between 
IRAS 23385+6053,
\MI\ and \MII\ are the lack of free-free emission towards the
core and the presence of a compact outflow driven by a
source embedded in the core. 

Given the strong similarities between \MI , \MII\ and IRAS 23385+6053,
it seems plausible that also the nature of the embedded YSOs is 
approximately the same.
Molinari et al. (\cite{mol98b}) propose two possible scenarios for
IRAS 23385+6053, that the object which lies 
inside the core is a B0 ZAMS star, and the \HII\ region is not detectable 
because it is very compact and optically thick; another possibility is that
there is no \HII\ region because
the central object is still in the proto-stellar phase. 
In both scenarios, the embedded source is young and massive,
while the stars of the surrounding cluster are more evolved. These 
scenarios are plausible also for \MI\ and \MII .
In the first hypothesis, the objects embedded inside I21307-C and I22172-C
are B1-B2 ZAMS stars; in the
second, they are intermediate-mass protostars, and 
the luminosity observed is due to accretion. 
Behrend \& Maeder (\cite{behrend}) computed evolutionary tracks for
protostars from 1 to 85 
$M_{\odot}$: in their models the luminosities of our sources correspond to 
protostellar masses of $\sim 5-8\;M_{\odot}$.
However, this value must be regarded with caution, because the authors
assume the relation found by Churchwell (\cite{church2}) between
$L$ and $\dot{M}_{\rm out}$, which in the high-mass 
regime does not provide a good fit to the data.

As discussed in Sect.~\ref{nature}, the protostellar 
hypothesis seems to be more likely. However,
to confirm this conclusion further interferometric observations
are needed. In particular, it would be important to measure the temperature
of the cores with great accuracy. In fact, 
newly formed massive stars lie inside ``hot cores'', namely compact molecular
regions with kinetic temperature higher than $\sim100$ K (Kurtz et al. 
\cite{kurtz}). Hence,  as discussed by Fontani et 
al. (\cite{fonta2}), measuring the core temperature would be an important
step to improve our understanding of their evolutionary stage.

\subsection{The formation of "Clusterinos"}
\label{clusterinos}

As we have discussed in the previous sections, it is very likely that
the outflows, the bulk of the bolometric luminosity and other properties of
the regions are dominated by the most massive forming star. Nevertheless,
our infrared images reveal that small young stellar aggregates are 
also found in the same regions (see also Molinari et al.~1998b
and Fontani et al.~2004). It is interesting to explore the 
idea that, along with the 5--8~M$_\odot$ protostar, a small stellar
cluster is forming within the molecular clump, a ``clusterino'' similar
to those found around the more evolved early type Herbig~Be stars
(see e.g. Testi et al.~\cite{Tea99}). The values of the clusterinos
richness indicators that we derived in Sect.~\ref{srclus} are relatively
low if compared with the values for Herbig~Be clusters. This is not 
unexpected, as our observations are not as sensitive (in terms of the least
massive objects detectable) as those of Testi et al.~(\cite{Tea98}),
this is mainly due to the larger distance of our sources compared with the
sample of Herbig~Be stars. An additional reason could be that our 
objects are generally younger and more extincted, and, finally, it is possible
that the clusterinos have not been completely assembled yet. 

From the photometric data alone, it is difficult to estimate the membership 
of individual objects (see also the discussion in the Testi et al. papers on 
Herbig Ae/Be clusterinos). 
Assuming that all the sources within 
$\sim$30$^{\prime\prime}$ from the IRAS nominal position are cluster members
and all the other sources in the field 
are not related to the clusterinos, it is possible to derive an estimate
of the most luminous (already formed) star and the integrated near infrared
emission for each clusterino. The result of this exercise shows that 
the most massive star in the \MI\ region is consistent with a moderately
reddened (A$_V\sim 4$mag) B6 ZAMS, while in \MII\ the brightest near
infrared star close to the IRAS position could be a more embedded
(A$_V\sim 9$mag) B1.5 ZAMS. The integrated near infrared fluxes at
$\sim$2.2~$\mu$m are $\sim$5~mJy and $\sim$90~mJy for \MI\ and \MII ,
respectively. These numbers show that the {\it already formed} members
of the clusterinos contribute only to a minimal fraction of the 
emission from the region in the case of \MI , while in \MII\ a 
substantial fraction of the total luminosity could be ascribed to the
most massive member.

It is important to point out that these estimates of the most massive
member of the clusterinos and their integrated emission at near infrared
wavelengths have been made assuming that all objects close to the IRAS source
are {\it bona fide} members. Clearly this is a simplifying assumption:
it is likely that an important contribution from foreground and background
objects is affecting our estimates. The comparison of the {\it richness
indicator} values and the number of objects close to the IRAS source shows
that this contamination could be as high as 50\%. Unfortunately, 
with our data alone, it is not
possible to pin point the contaminating objects, hence our estimates
should be regarded as upper limits (both for the luminosity of the most massive
star and the integrated flux of the clusterinos).

\section{Summary and conclusions}
\label{conc}

We have observed two massive proto-stellar candidates, IRAS 21307+5049 and
IRAS 22172+5549, in the \CO\ (1--0) and (2--1), 
\CII\ (2--1), \HCOpI\ (1--0), \ace\ (6--5), (8--7) and (13--12) molecular
lines, and in the continuum at various wavelengths from the near-infrared
to the millimetric range. The following
results have been obtained:
\begin{itemize}
\item Maps of the 3~mm continuum observed with the OVRO
interferometer reveal in both sources the presence of a compact, massive core, 
whose linear diameter and mass are $\sim 0.09$ pc and $\sim 50\;M_{\odot}$ 
for \MI , and $\sim 0.04$ pc and $\sim 40\;M_{\odot}$ for \MII ;
\item Near-infrared images at 1.7 and 2.2 $\mu$m indicate that clusters
of more evolved stars surround both cores. The continuum maps and the SEDs 
suggest that for both sources the core
is mainly responsible for the emission at wavelengths longer than 
$\sim25\;\mu$m, while the emission at shorter wavelengths is due
to the surrounding cluster. The SEDs indicate that the core luminosity 
is a few $10^{3}\;L_{\odot}$ for both sources.
\item By mapping the \CO\ (1--0) and (2--1) line wings, we find  
outflows associated with both sources that are driven by YSOs embedded
in the cores. Outflows have
masses of a few to a few tens of $M_{\odot}$, dynamical timescales of
$\sim10^{4}-10^{5}$ yr and mass loss rates of $\sim$ a few times 
$\sim10^{-4}M_{\odot}{\rm yr^{-1}}$.
Images of the H$_{2}$ and [FeII] lines at
near-infrared wavelengths suggest the presence of an ionized jet in the
inner part of the outflow associated with \MII .
\item The flows physical parameters of both sources are consistent with 
those of a sample of massive protostar candidates 
(Beuther et al. \cite{beuther}) with comparable luminosity. 
The values of the accretion efficiencies, of $\sim10^{-2}$, are 
consistent with values derived assuming that in these sources the luminosity
is accretion-dominated. 
\item Based on the similarity between our sources and the massive 
protostar candidate IRAS 23395+6053, we believe that the massive YSOs
embedded inside the cores are protostars accreting material from
the environment, the masses of which at present are  
$\sim 5-8\;M_{\odot}$, assuming that the luminosity observed is due to 
accretion onto the forming star.
\item Near-infrared images suggest that, together with the massive
protostar, a small stellar cluster (``clusterino'') is forming inside the 
molecular clump, similar to those found around Herbig Be stars. 
Assuming that the sources within 30\asec\ from the IRAS position are all
{\it bona fide} members of the clusterino, we have derived that
the already formed stars of the
clusterino contribute only to a small fraction to the total observed
luminosity in \MI , while in \MII\ a significative fraction of the total
luminosity could be due to the most massive members.
\end{itemize}
\begin{acknowledgements}
We thank Hans Ungerechts for his precious help with the
IRAM-30m observations, and Henrik Beuther for stimulating disussions.
This publication makes use of data products from the Two Micron All Sky Survey,
which is a joint project of the University of Massachusetts and the Infrared
Processing and Analysis Center/California Institute of Technology, funded by
the National Aeronautics and Space Administration and the National Science
Foundation.
The OVRO mm array is supported by NSF grant
AST-99-81546, research on young stars and disks
is also supported by the {\it Norris Planetary Origins Project} and
NASA {\it Origins of Solar Systems} program (grant NAG5--9530).
This paper is partly based on observations made with the Italian
Telescopio Nazionale Galileo (TNG) operated on the island of La
Palma by the Centro Galileo Galilei of the INAF (Istituto Nazionale
di Astrofisica) at the Spanish Observatorio del Roque de los Muchachos
of the Instituto de Astrofisica de Canarias.
The NICS/TNG observations were performed in service mode by
the TNG staff, we especially acknowledge the help of
Antonio Magazz\`u.
We also thank Marina Cecere for help with the Palomar observations and
data reduction. 

\end{acknowledgements}
{}

\end{document}